\documentclass[10pt,journal,compsoc]{IEEEtran}

\ifCLASSOPTIONcompsoc
  \usepackage[nocompress]{cite}
\else
  \usepackage{cite}
\fi

\ifCLASSINFOpdf
\else
\fi

\usepackage{balance}
\usepackage[ruled,noend]{algorithm2e}
\usepackage{booktabs} 
\usepackage{multirow}
\usepackage{array}
\usepackage{amsmath,amssymb,amsfonts}
\usepackage{graphicx}
\usepackage{textcomp}
\usepackage{soul}
\usepackage[labelformat=simple]{subcaption}

\usepackage{soul, xcolor}

\begin{document}

\title{Covert Channel-Based Transmitter Authentication in Controller Area Networks}
%
%
%
%

\author{Xuhang~Ying,~\IEEEmembership{Member,~IEEE,}
        Giuseppe~Bernieri,~\IEEEmembership{Member,~IEEE,}
        Mauro~Conti,~\IEEEmembership{Senior~Member,~IEEE,}
        Linda~Bushnell,~\IEEEmembership{Fellow,~IEEE,}
        and~Radha~Poovendran,~\IEEEmembership{Fellow,~IEEE}
\IEEEcompsocitemizethanks{\IEEEcompsocthanksitem X. Ying, L. Bushnell, and R. Poovendran are with the Department of Electrical and Computer Engineering, University of Washington, Seattle, WA, 98195. 
E-mail: \{xhying, lb2, rp3\}@uw.edu
\IEEEcompsocthanksitem G. Bernieri and M. Conti are with the Department of Mathematics, University of Padua, Padua, Italy. 
Email: \{bernieri, conti\}@math.unipd.it
\IEEEcompsocthanksitem 
This work was supported in part by NSF grant CNS-1446866 under the CPS program, ONR grants N00014-16-1-2710 and N00014-17-1-2946, and ARO grant W911NF-16-1-0485. This work was also supported by a grant of the Italian Presidency of the Council of Ministers.
\IEEEcompsocthanksitem Part of this work was presented at ACM/IEEE ICCPS 2019 \cite{ying2019tacan}.
}
\thanks{Manuscript received April XX, 2019; revised August XX, 2019.}}

%
%

\markboth{IEEE Transactions on Dependable and Secure Computing, ~Vol.~XX, No.~XX, August~2019}%
{Ying \MakeLowercase{\textit{et al.}}: TACAN: Transmitter Authentication through Covert Channels in Controller Area Networks}
%


\IEEEtitleabstractindextext{%

\begin{abstract}
In recent years, the security of automotive Cyber-Physical Systems (CPSs) is facing 
urgent threats due to the widespread use of legacy in-vehicle communication systems. 
As a representative legacy bus system, the Controller Area Network (CAN) hosts
Electronic Control Units (ECUs) that are crucial for the vehicles functioning. 
In this scenario, malicious actors can exploit the CAN vulnerabilities, such as the lack of built-in authentication and encryption schemes, to launch CAN bus attacks (e.g., suspension, injection, and masquerade attacks) with life-threatening consequences (e.g., disabling brakes).
In this paper, we present TACAN (Transmitter Authentication in CAN), which provides secure authentication of ECUs on the legacy CAN bus by exploiting the \textit{covert channels}, without introducing CAN protocol modifications or traffic overheads (no extra bits or CAN messages are used). 
TACAN turns upside-down the originally malicious concept of covert channels and exploits it to build an effective defensive technique that facilitates transmitter authentication via a centralized, trusted Monitor Node.
TACAN consists of three different covert channels for ECU authentication:
1) the Inter-Arrival Time (IAT)-based, leveraging the IATs of CAN messages; 2) the Least Significant Bit (LSB)-based, concealing authentication messages into the LSBs of normal CAN data; and 3) a hybrid covert channel, exploiting the combination of the first two.
In order to validate TACAN, we implement the covert channels on the University of Washington (UW) EcoCAR (Chevrolet Camaro 2016) testbed.
We further evaluate the bit error, throughput, and detection performance of TACAN through extensive experiments using the EcoCAR testbed and a publicly available dataset collected from Toyota Camry 2010. 
We demonstrate the feasibility of TACAN and the effectiveness of detecting CAN bus attacks, highlighting no traffic overheads and attesting the regular functionality of ECUs.

\end{abstract}

}

\maketitle

\IEEEdisplaynontitleabstractindextext

%
\IEEEpeerreviewmaketitle

\IEEEraisesectionheading{\section{Introduction}}
\label{sec:introduction}

\IEEEPARstart{N}{owadays}, the technological evolution enables an increasingly more invasive interconnection between automobiles and digital devices.
Automotive manufacturers have developed a variety of innovative features such as the smart transportation data assistants by leveraging network connectivity such as the vehicle-to-vehicle communication.
While these features greatly improve the customer experience, Electronic Control Units (ECUs) that are externally accessible provide an entry point for an adversary to infiltrate the originally isolated in-vehicle communication network, notably the Controller Area Network (CAN) \cite{ISO2015CAN,bosch1991CAN}.
Since the CAN bus is a broadcast medium without authentication, a compromised ECU can masquerade as any targeted ECU by transmitting messages with the forged message ID (masquerade attack~\cite{checkoway2011comprehensive}).
Modern externally accessible ECUs
with additional connectivity interfaces such as cellular, Wi-Fi or Bluetooth 
disrupt the closed in-vehicle network assumption.
Consequently, the CAN bus is vulnerable to cyber attacks, such as disabled brakes~\cite{checkoway2011comprehensive} and remotely controlled steering~\cite{tesla2016remote}.
Despite of the well-known security vulnerabilities of the CAN bus, its widespread use imposes an urgent need for security solutions that  guarantee the functionality and safety of today's automobiles and future's autonomous cars~\cite{bojarski2016end,wyglinski2013security}.

The use of cryptographic primitives, such as message authentication, represents one possible way to defend against CAN bus attacks (notably the masquerade attack).
However, due to the low throughput and tight bit budget of the CAN protocol, it is challenging to deploy cryptographic schemes in practice and current solutions, such as~\cite{herrewege2011canauth,hazem2012lcap,kurachi2014cacan,radu2016leia}, would require protocol modifications or introduce traffic overheads.
An alternative is to deploy anomaly-based Intrusion Detection Systems (IDSs) without modifying the CAN protocol~\cite{cho2016finger,muter2011entropy,cho2017viden,choi2018voltageids}, including timing-based and voltage-based IDSs.
The timing-based IDS in~\cite{cho2016finger} exploits CAN message periodicity to estimate clock skew as a unique fingerprint to detect masquerade attacks. 
Nevertheless, it was later shown to be ineffective against the cloaking attack that modifies the inter-transmission time to emulate the clock skew of the targeted ECU~\cite{sagong2018cloaking,ying2018shape}. 
The voltage-based IDSs~\cite{murvay2014source,cho2017viden,choi2018voltageids,kneib2018scission} attempt to 
fingerprint the attacker through voltage signal characteristics. 
However, if the attacker uses IDs that the compromised ECU is allowed to use under normal conditions, the attack will not be detected.


\begin{figure}[t!]
\centering
\includegraphics[width=1\columnwidth]{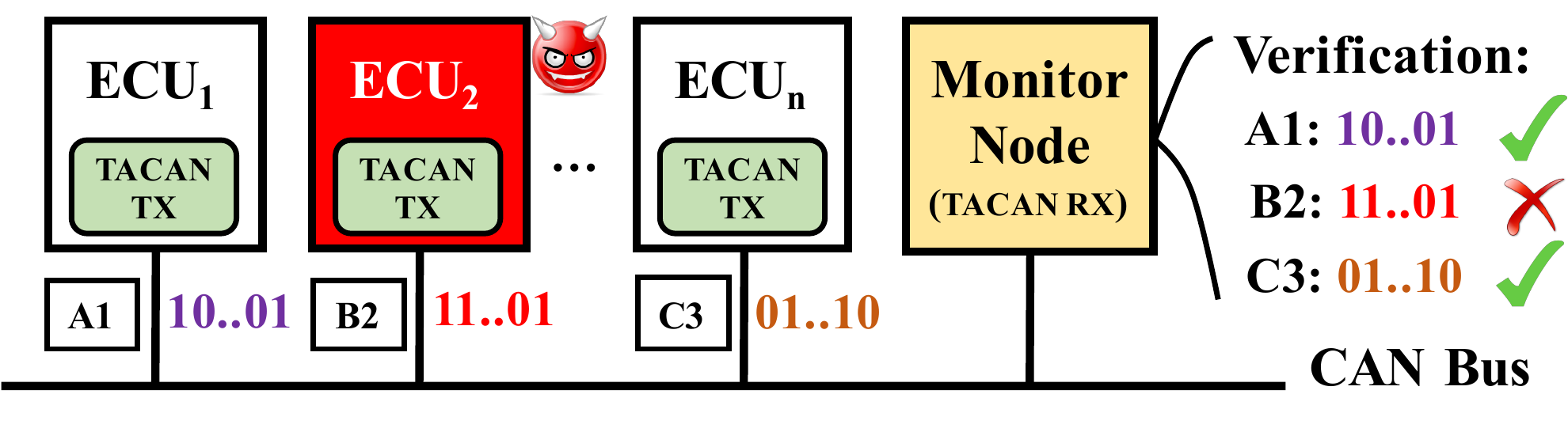}
\caption{Illustration of TACAN. Legitimate ECUs transmit unique authentication messages that are embedded into the timing and LSBs of normal CAN messages (e.g., A1, B2, and C3) using the proposed covert channel methodologies. 
The Monitor Node authenticates transmitting ECUs by verifying the received authentication messages. If the compromised ECUs cannot generate valid authentication messages, then the attack will be detected by the Monitor Node.}
\label{fig:main_idea}
\end{figure}

In this work, we present TACAN, a novel security framework that allows a centralized, trusted \textit{Monitor Node} (MN) to verify the authenticity of a transmitting ECU and detect CAN bus anomalies. Fig.~\ref{fig:main_idea} highlights the main idea of our TACAN framework.
In TACAN, a master key is shared between an ECU and the MN for generating shared session keys. 
Consistently with \cite{hazem2012lcap,radu2016leia, herrewege2011canauth}, we assume that the keys are stored in the tamper-resistant memory of a security module such as the Trusted Platform Module (TPM) \cite{TPMautomotive}. 
%
%
%
%
The ECU embeds unique authentication messages into CAN messages and continuously transmits them through covert channels, which can be received and verified by the MN.

Therefore, if the attacker has no access to the TPM of the targeted ECU, 
it cannot use the compromised ECU or external device to generate valid authentication messages, thus causing verification failures and triggering the alarm at the MN side.
Moreover, CAN bus attacks, such as the suspension and injection attacks, that interrupt the transmission of normal CAN messages with embedded authentication messages will cause continuous authentication message loss or reception failures, and therefore they will be detected by TACAN.
The main benefits of using covert channels for TACAN are that they do not introduce protocol modifications or traffic overheads (i.e., extra bits or messages).
In addition, by requiring ECUs to transmit authentication messages much less frequently than per-message authentication schemes, TACAN can significantly reduce the computational burden of the resource-constrained ECUs.


\vspace{0.1cm}
\noindent \textbf{Contributions: }
In this paper, we make the following contributions: 
\begin{itemize}
    \item We identify and exploit covert channels to facilitate ECU authentication on the CAN bus. 
    Hence, covert channels are used for security instead of malicious communication.
    
    
    \item We propose TACAN and three novel covert channels for authentication message transmission: 1) the IAT (Inter-Arrival Time)-based covert channel that modifies the inter-transmission times (ITTs) of normal CAN messages to affect the IATs observed by the MN;
    2) the LSB (Least Significant Bit)-based covert channel that hides the authentication bits in the LSBs of the data payload of normal CAN messages; 
    3) an hybrid covert channel that combines the first two covert channels.
    
    \item For the IAT-based covert channel, we analyze and model the bit error probability as a function of covert channel parameters. We also study its impact on the CAN bus schedulability in terms of the worst-case response time.

    \item We validate TACAN and the covert channels on the UW EcoCAR~\cite{ecocar}. 
    We also conduct extensive experiments to evaluate the performance of TACAN using the EcoCAR testbed as well as the publicly available Toyota dataset~\cite{toyota2010dataset}.
    Our results show that with a properly configured IAT-based covert channel, the experimental bit error probabilities are within 0.3\%. 
    As for the LSB-based covert channel, its bit error probability is equal to that of a typical CAN bus, which is $3.1\times 10^{-7}$\%. 
    The hybrid covert channel generally leads to higher channel and authentication throughput.
    We also show that our TACAN-based detector can detect CAN bus attacks with a very high detection probability, while keeping the false alarm probability less than 0.33\% by setting the detection threshold to $2$ or higher.

\end{itemize}

\noindent \textbf{Organization.}
The remainder of this paper is organized as follows.
Section~\ref{sec:related_work} reviews the related work.
Section~\ref{sec:system_and_adversary_model} presents our system and adversary models.
Section~\ref{sec:proposed_scheme} presents TACAN. 
Section~\ref{sec:evaluation} presents experimental evaluation.
Section~\ref{sec:conclusion} concludes this paper. 
\section{Related Work}
\label{sec:related_work}
Recent experimental studies have demonstrated that an attacker is able to infiltrate in-vehicle ECUs physically or remotely and mount cyber attacks that would cause potentially life-threatening consequences by disabling brakes or overriding steering~\cite{miller2015remote,checkoway2011comprehensive}.
One way to secure the CAN bus is to deploy anomaly-based IDSs based on traffic analysis (e.g., timing/frequency~\cite{hoppe2008security}), 
entropy~\cite{muter2011entropy}, or physical invariants such as clock skew~\cite{cho2016finger} and voltage signal characteristics~\cite{murvay2014source,cho2017viden,choi2018voltageids,kneib2018scission}. 
For clock skew-based IDSs, it has been shown in \cite{sagong2018cloaking,ying2018shape} that the adversary can effectively manipulate the timing of transmitted messages to emulate the clock skew of the targeted ECU and evade the detection.
While voltage-based IDSs are effective against ongoing masquerade attackers, they cannot detect a compromised ECU before attacks are launched (e.g., a stealthy attacker may not launch the attack until the car is in drive mode). 
In addition, it has been recently shown in \cite{sagong2018exploring} that the extra wires required by voltage-based IDSs may introduce new attack surfaces.

Researchers have also proposed to add cryptographic primitives such as Message Authentication Code (MAC) to the CAN bus, including CANAuth~\cite{herrewege2011canauth}, LCAP~\cite{hazem2012lcap}, CaCAN~\cite{kurachi2014cacan}, and LeiA~\cite{radu2016leia}. 
Nevertheless, the deployment of the above schemes faces several practical challenges. First, the CAN protocol has a tight bit budget for each CAN frame (up to 8 bytes for payload) and a low bus speed (typically 500 kbps). 
As a result, authentication information will have to consume space in the CAN message such as the ID or data field or introduce additional CAN messages, which leads to traffic overheads or an increase in the bus load~\cite{hazem2012lcap,kurachi2014cacan,radu2016leia}.
Second, it is also computationally expensive for resource-constrained ECUs to perform cryptographic calculations for each message.
In this work, we focus on transmitter authentication by having each ECU transmit unique authentication messages as its digital fingerprint.
The main novelty of this work is the use of covert channels, a well-known malicious technique that is converted into defensive applications for authentication purposes.
By leveraging covert channels as out-of-band channels, our scheme can avoid traffic overheads without requiring protocol modifications.

In literature, a covert channel refers to a type of cyber-attack that maliciously transfers information between two possibly malicious entities by exploiting the communication channels that are not intended for information transfer. 
It has been widely studied in computer network protocols \cite{zander2007survey}.
Broadly speaking, there are two categories of covert channels: timing-based and storage-based. 
In timing-based covert channels, only the timing of events or traffic is modified to transfer information but the data contents remain intact.
The storage-based covert channels hide data in a shared resource (e.g., a storage location). 

Recently, researchers start exploring covert channels in the context of embedded networks.
In~\cite{taylor2017enhancing}, Taylor \textit{et al.} discussed the use of covert channels for integrity check for the Modbus/TCP protocol used by industrial control system applications.
In \cite{groza2018incanta}, Groza \textit{et al.} proposed a time-covert authentication scheme for the CAN protocol, which uses fine-grained timing control to embed authentication information in clock skews.
However, timing control in the order of tens of nanoseconds is challenging in practice, and the proposed scheme is very sensitive to message priority and the traffic. 
In addition, since the clock skew needs to be estimated from arrival times of many CAN messages, it seems to contradict with the proposed scheme that performs authentication for each CAN message.

In this paper, we develop three practical covert channels for transmitter authentication on the legacy CAN bus, based on the periodicity of CAN messages and bit presentation of floating sensor values. We also evaluate the proposed covert channels using CAN traffic data collected from real vehicles.
\section{System and Adversary Models}
\label{sec:system_and_adversary_model}
In this section, we present the system model (Section~\ref{sec:system_model}) and the adversary model (Section~\ref{sec:adversary_model}) for the CAN bus.
A list of frequently used notations is provided in Table~\ref{table:notation}.

\begin{table}[t!]
	\centering
	\caption{Frequently used notations.
	}
	\begin{tabular}{|c|l|}
		\hline
		\textbf{Notation} & \textbf{Description} \\
		\hline
		$MK^{(k)}$ & Master key (MK)  for message ID $k$ \\ \hline
		$SK^{(k)}$ & Session key (SK) for message ID $k$ \\ \hline
		$C_{s}^{(k)}$ & Session key counter for message ID $k$ \\ \hline
		$C_{m}^{(k)}$ & Authentication message counter for message ID $k$\\ \hline
		$T$ & CAN message period (sec)\\ \hline
		$S$ & Clock skew (ppm) \\ \hline
		$t_i$ & Transmit time of the $i$-the CAN message \\ \hline
		$a_i$ & Arrival time of the $i$-the CAN message \\ \hline
		$\eta_i$ & Noise in the arrival time of the $i$-the CAN message \\ \hline
		\multirow{ 2}{*}{$\Delta t_i$} & Inter-transmission time (ITT) between the $(i-1)$-th \\ & and the $i$-th messages \\ \hline
		\multirow{ 2}{*}{$\Delta a_i$} & Inter-arrival time (IAT) between the $(i-1)$-th \\& and the $i$-th messages \\ \hline
		$\Delta \bar{a}[i]$ & The $i$-th sample of averaged IATs\\ \hline 
		$\delta$ & Deviation (added to ITTs at the transmitter side) \\ \hline
		$L$ & Window length or number of least significant bits \\ \hline
		$\mu, \sigma$ & Mean and standard deviation of IATs \\ \hline 
		
	\end{tabular}
	
	\label{table:notation}
	\normalsize
\end{table}

\subsection{System Model}
\label{sec:system_model}

\begin{figure}[ht!]
\centering
\includegraphics[width=1\columnwidth]{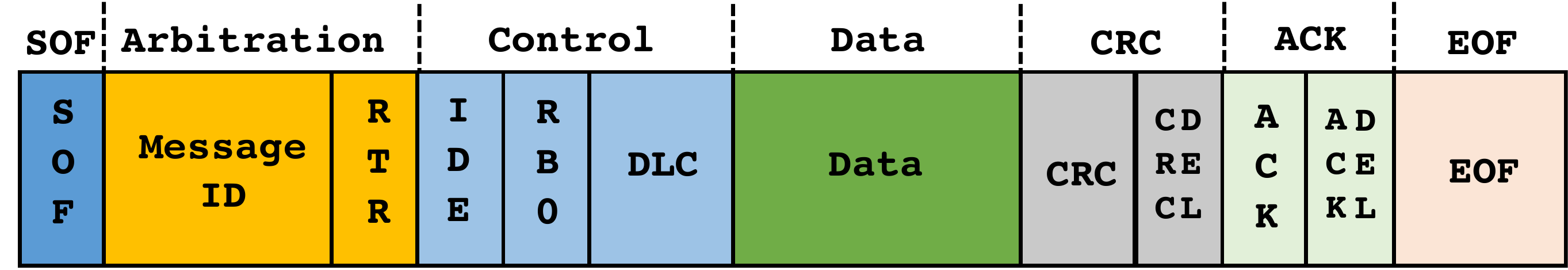}
\caption{Illustration of CAN frame structure.}
\label{fig:can_frame_structure}
\end{figure}

\noindent \textbf{CAN bus.}
The CAN bus is a broadcast medium that allows all connected ECUs to communicate with each other and observe all CAN message transmissions.
As shown in Fig.~\ref{fig:can_frame_structure}, each CAN frame or message has a set of predefined fields, including the Start of Frame (SOF) field, the Arbitration field (including a 11-bit message ID for the base frame format or a 29-bit message ID for the extended frame format), the Control field, the Data field (up to 8 bytes), the Cyclic Redundancy Check (CRC) field, the Acknowlegement (ACK) field, and the End of Frame (EOF) field.

When two (or more) ECUs attempt to transmit messages at the same time, a procedure based on priority (a smaller message ID indicates a higher priority) called arbitration is used to determine the winner. 
Besides, CAN messages do not have transmit timestamps and do not support encryption or authentication. 

\vspace{0.1cm}
\noindent \textbf{Clock skew.} 
On automotive CAN buses, the majority of CAN messages are  transmitted periodically as per ECUs' local clocks\footnote{In the UW EcoCAR (Chevrolet Camaro 2016) \cite{ecocar}, all of the 89 messages with distinct IDs are periodic with periods ranging from 10 ms to 5 sec. In the Toyota Camry 2010 \cite{toyota2010dataset}, 39 of 43 messages can be considered periodic. In the Dodge Ram Pickup 2010 in \cite{cho2016finger}, all of the 55 distinct messages are periodic. While CAN message periodicity depends on the manufacturer and the model, the above examples suggest that periodic CAN messages are very common and even dominant on the CAN bus of commercial automobiles.}, and there is no clock synchronization in CAN.
Hence, the frequencies of local clocks are different, as captured by the notion of clock skew -- a physical property caused by variations in the clock's hardware crystal. 

Let $\mathcal{C}_A(t)$ be the time reported by clock $A$ and $\mathcal{C}_{true}(t)=t$ be the true time. 
According to the Network Time Protocol (NTP)~\cite{mills1992NTP}, the clock offset of clock A is defined as $O_A(t)=\mathcal{C}_A(t)-\mathcal{C}_{true}(t)$, 
and the clock skew is the first derivative of clock offset, i.e., $S_A(t)=O'_A(t)=\mathcal{C}'_A(t)-1$, which is measured in microseconds per second ($\mu$s/s) or parts per million (ppm).
In the absence of a true clock, the relative clock offset and relative clock skew can be defined with respect to a reference clock.
Throughout this paper, we consider the receiving ECU's clock as the reference clock. Hence, we refer to the relative clock offset and the relative clock skew as clock offset and clock skew, respectively.

\begin{figure}[t!]
\centering
\includegraphics[trim=0cm 0.7cm 0cm 0cm, clip=true, width=1\columnwidth]{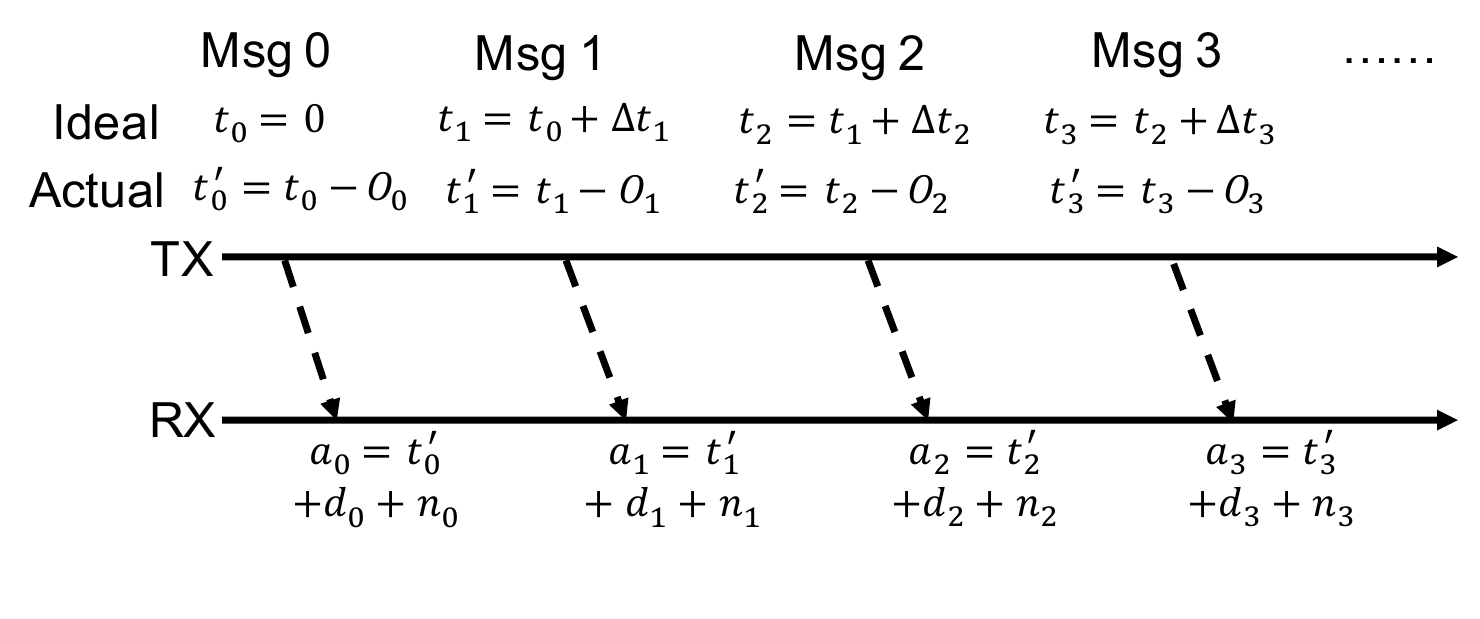}
\caption{Illustration of our timing model. 
Due to the clock skew of the transmitter's clock, the actual transmit time of message $i$ is $t_i'=t_i-O_i$ as per the receiver's clock, where $O_i$ is the accumulated clock offset up to message $i$. 
}
\label{fig:timing_model}
\end{figure}

\vspace{0.1cm}
\noindent \textbf{Timing model.}
As shown in Fig.~\ref{fig:timing_model}, we let $t_i$ be the transmit time of message $i$ (assuming $t_0=0$) and $\Delta t_i = t_i - t_{i-1}$ be the inter-transmission time (ITT) according to the transmitter's clock.
If messages are periodically transmitted every $T$ seconds, we have $\Delta t_i=T$ and $t_i=iT$. 
In the ideal case where the transmitter's clock is synchronized with the receiver's clock, we have $t'_i=t_i$, where $t'_i$ is the transmit time according to the receiver's clock. 
Nevertheless, in practice, there exists a clock skew in the transmitter's clock relative to the reference clock, which introduces an offset $O_i$ between the two clocks. 
Hence, the actual transmit time is $t'_i = t_i - O_i$ according to the reference clock. 

While the clock skew may be slowly varying due to factors including the temperature, it is almost constant over a short duration.
Given a clock skew $S$, the relationship between the elapsed time $t_i$ in the transmitter's clock and the elapsed time $t'_i$ in the receiver's clock is $S = (t_i - t'_i)/t'_i$.
Hence, we have $t_i' = t_i/(1+S)$, and $O_i = t_i - t_i' = \frac{S}{1+S} t_i$.
To capture random jitters, we assume $O_i = \frac{S}{1+S} t_i + \epsilon_i$, 
where $\epsilon_i$'s are i.i.d. zero-mean random variables.

After a random network delay of $d_i$ (due to message transmission, propagation, arbitration, and reception) and the zero-mean quantization noise $q_i$~\cite{zander2008measurement}, the arrival time of message $i$ is
\begin{align}
    a_i &= t_i - O_i + d_i + q_i \nonumber\\
    &=t_i - \frac{S}{1+S} t_i  + \eta_i =\frac{1}{1+S}t_i + \eta_i, \label{eq:arrival_timestamp}
\end{align}
where $\eta_i = -\epsilon_i + d_i + q_i$ is the total noise in the arrival timestamp. 
Since periodic CAN messages have the same message ID and  data length over time, it is reasonable to assume constant-mean network delays, i.e., $\mathbb{E}[d_i] = d$. 
Hence, $\eta_i$'s can be modeled as i.i.d. random variables with a mean of $d$ and a variance of $\sigma_\eta^2$. 

Denote the inter-arrival time (IAT) between messages $(i-1)$ and $i$ as $\Delta a_i$, which is given by
\begin{equation}
    \Delta a_i = a_i - a_{i-1} = \frac{1}{1+S} \Delta t_i + n_i, \label{eq:inter_arrival_time}
\end{equation}
where $n_i=\eta_i - \eta_{i-1}$ is the noise term.
For messages that are transmitted every $T$ seconds, we have $\Delta t_i = T$. 
Hence, the IATs have a mean of
\begin{equation}
    \mu \triangleq \mathbb{E}[\Delta a_i] = \frac{T}{1+S} \label{eq:IAT_mean}
\end{equation}
and a variance of 
\begin{equation}
    \sigma^2 \triangleq \text{Var}[\Delta a_i] = 2\sigma_\eta^2. \label{eq:noise_variance}
\end{equation}
When the clock skew is small (in the order of 100s of ppm), the impact of clock skew is negligible, and we have $\mu \approx T$.

\subsection{Adversary Model}
\label{sec:adversary_model}
We consider an adversary who attempts to infiltrate the CAN bus and launch stealthy attacks without being detected.
We assume that the adversary can passively monitor the CAN bus and observe all ongoing CAN transmissions. 
It has full knowledge of the deployed covert channels and thus can observe all authentication messages that are being transmitted.
In practice, there are usually two ways of gaining unauthorized access to the CAN bus: 1) compromise an in-vehicle ECU physically or remotely~\cite{checkoway2011comprehensive}, or 2) connect external device (a malicious ECU) to the CAN bus~\cite{koscher2010experimental}.
We assume that the adversary has no access to the keys stored in the TPM of the compromised and other legitimate ECUs.

\begin{figure}[t!]
    \centering
    \begin{subfigure}[b]{1\columnwidth}
        \includegraphics[width=\textwidth]{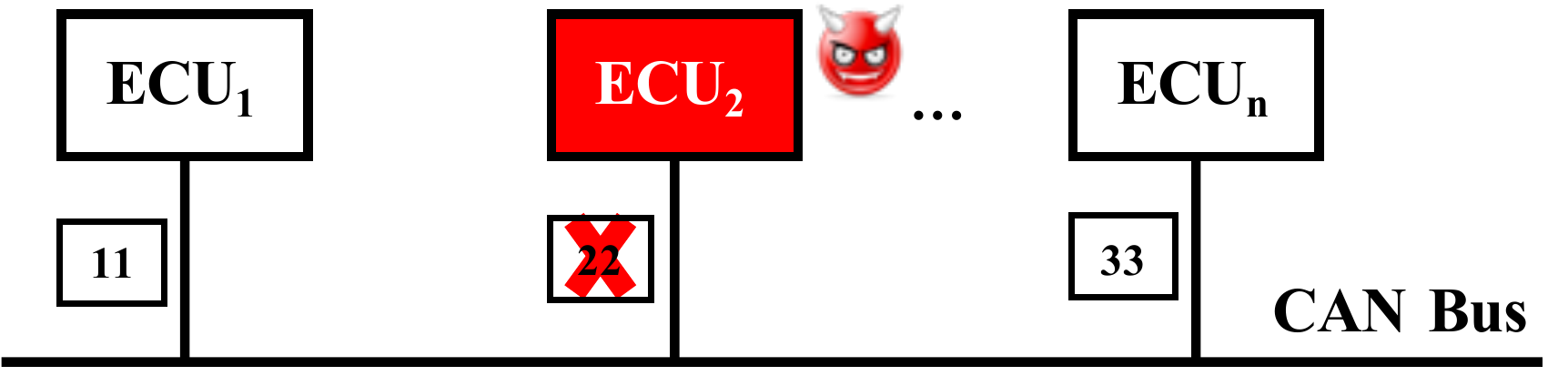}
        \caption{Suspension attack}
        \label{fig:attack_suspension}
    \end{subfigure}
    
    \vspace{0.2cm}
    
    \begin{subfigure}[b]{1\columnwidth}
        \includegraphics[width=\textwidth]{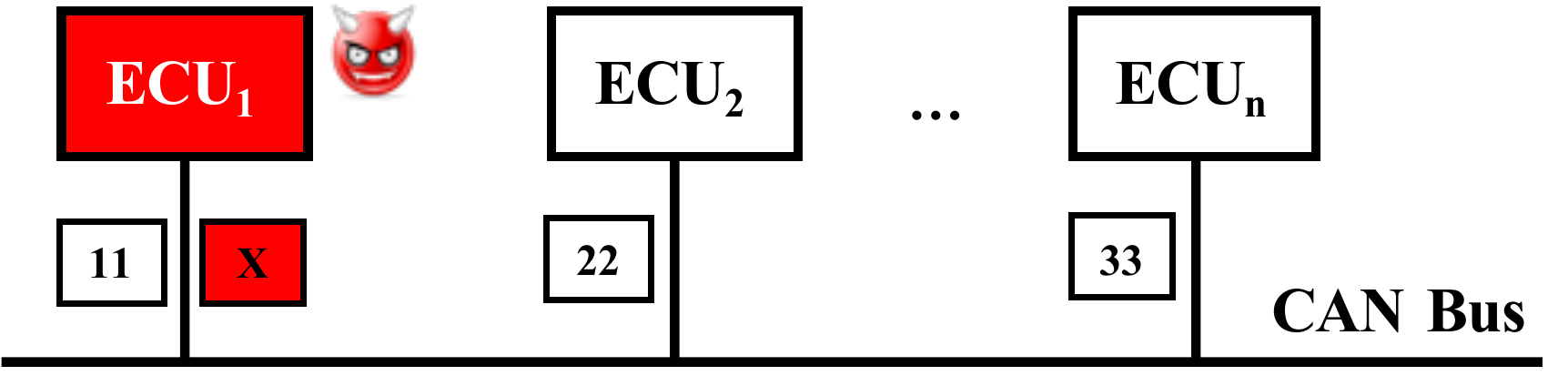}
        \caption{Injection attack}
        \label{fig:attack_fabrication}
    \end{subfigure}
    
    \vspace{0.2cm}
    
    \begin{subfigure}[b]{1\columnwidth}
        \includegraphics[width=\textwidth]{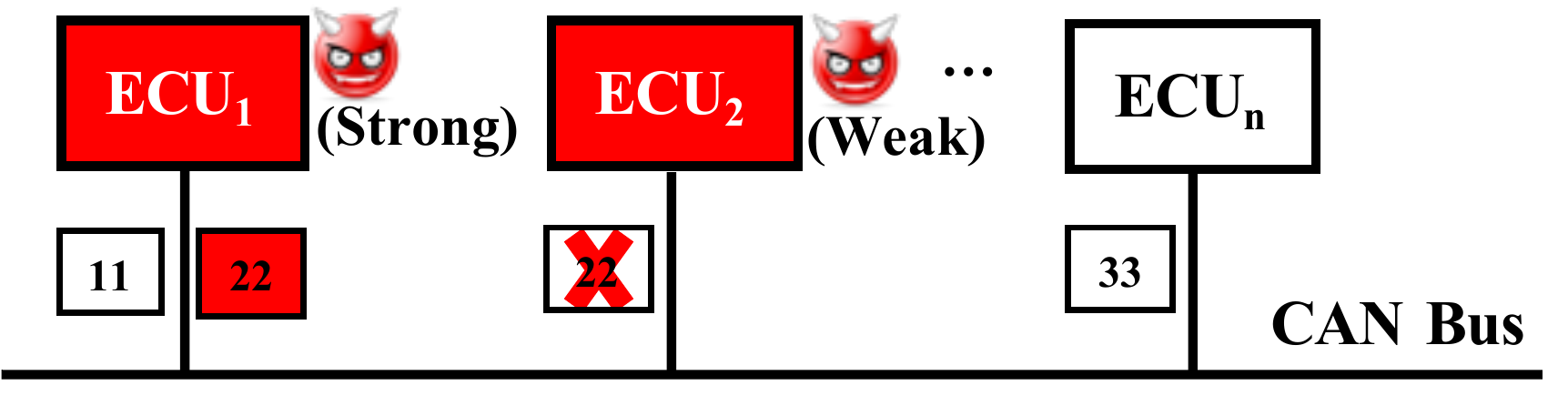}
        \caption{Masquerade attack}
        \label{fig:attack_masquerade}
    \end{subfigure}
    
    \caption{Three representative attacks on the CAN bus. (a) In the suspension attack, a legitimate message is prevented from being transmitted. 
    (b) In the fabrication attack, the adversary can use the compromised ECU to inject arbitrary CAN messages. 
    (c) In the masquerade attack, the adversary stops $\text{ECU}_2$ from transmitting messages with ID=0x22 and uses $\text{ECU}_1$ to transmit messages with ID=0x22.
    }
    \label{fig:attack_model}
\end{figure}

The adversary can use the compromised or malicious ECU to perform three representative attacks: 1) suspension attack, 2) injection attack, and 3) masquerade attack, as considered in~\cite{lin2012cyber,cho2016finger,sagong2018cloaking}.
As illustrated in Fig.~\ref{fig:attack_suspension}, a suspension attacker prevents the compromised ECU from transmitting certain messages, whereas an injection attacker fabricates and injects CAN messages of arbitrary choices of message ID, content, and timing, as sketched in Fig.~\ref{fig:attack_fabrication}.
Injection attacks can lead to more sophisticated attacks such as the DoS 
attack~\cite{hoppe2008security} and the bus-off attack~\cite{cho2016error}. 
In the masquerade attack, the adversary will need to compromise two ECUs -- one is weakly compromised and acts as the weak attacker who can only launch suspension attacks, whereas the other one is fully compromised and acts as the strong attacker who can launch both suspension and injection attacks. 
In the example in Fig.~\ref{fig:attack_masquerade}, the adversary suspends the weakly compromised $\text{ECU}_2$ from transmitting messages with ID=0x22 and uses the fully compromised $\text{ECU}_1$ to inject messages with ID=0x22 claiming to originate from $\text{ECU}_2$. 
Compared to the suspension and injection attacks, the masquerade attack is stealthier and thus more difficult to detect. 

\section{TACAN}
\label{sec:proposed_scheme}
In this section, we present the architecture (Section~\ref{sec:tacan_architecture}) and the transmitter authentication protocol of TACAN (Section~\ref{sec:transmission_authentication_protocol}).
We then present three covert channels for transmitting authentication messages: 
1) IAT-based (Section~\ref{sec:IAT_based_covert_channel}), 2) LSB-based (Section~\ref{sec:lsb_based}), and 3) hybrid  (Section~\ref{sec:hybrid}).

\begin{figure*}[t!]
\centering
\includegraphics[width=.8\textwidth]{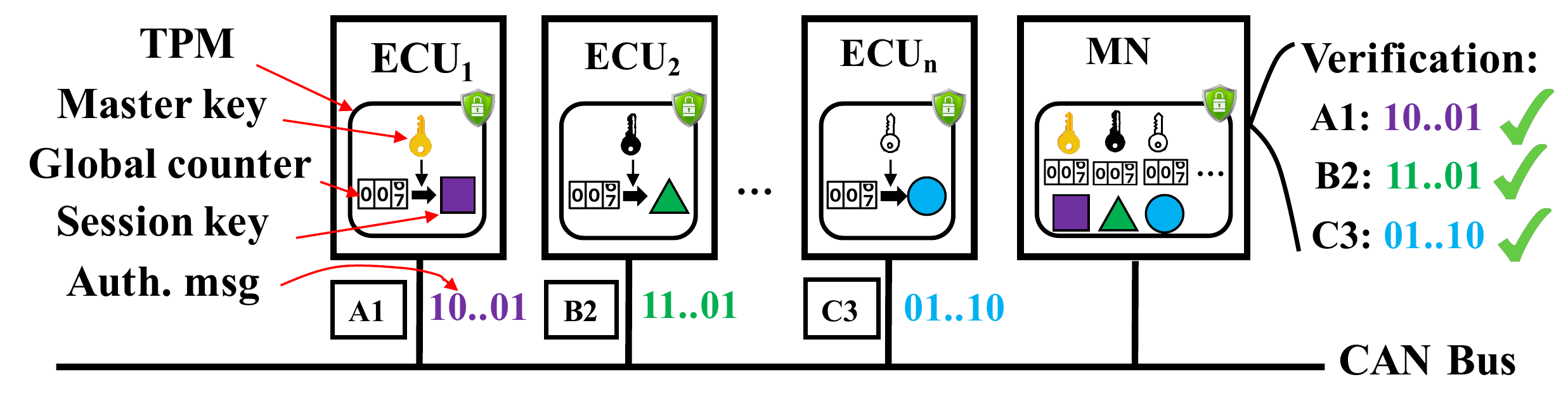}
\caption{Illustration of the architecture and the transmitter authentication protocol of TACAN. A master key is shared and a session key counter is synchronized between each ECU and the MN. The ECU and the MN generate the same session key from the master key and the session key counter, and they use it to generate and verify authentication messages, respectively. Unsuccessful reception and verification of authentication messages implies a possible CAN bus anomaly. 
The master key, the session key counter, and the generated session keys are stored securely in the TPM of the ECU and the MN.
}
\label{fig:TACAN_arch}
\end{figure*}

\subsection{TACAN architecture}
\label{sec:tacan_architecture}
As illustrated in Fig.~\ref{fig:TACAN_arch}, TACAN consists of in-vehicle ECUs and a centralized, trusted Monitor Node (MN).
The MN is installed by the manufacturer during production and requires direct physical access by authorized parties (e.g., an authorized repairs shop) to prevent potential tampering and compromises.
The deployed covert channels are configured during production or re-configured during maintenance to ensure successful establishment of one-way communication of authentication information from ECUs to the MN.

Similar to~\cite{hazem2012lcap,radu2016leia,herrewege2011canauth}, we assume that a master key (MK) is pre-shared between each ECU and the MN, and it is stored in the TPM. 
Updating of MKs (e.g., when adding or replacing an ECU) should again require direct physical access by authorized parties to the involved ECUs.
The procedure of key updating is outside the scope of this paper.
During operation, the ECU and the MN will generate the same session key (SK) from the MK and the synchronized SK  counter, and further use it to generate and verify authentication messages, respectively.
We now describe the transmitter authentication protocol in more detail.


\subsection{Transmitter Authentication Protocol}
\label{sec:transmission_authentication_protocol}
Inspired by the work in~\cite{radu2016leia}, the MN in TACAN performs unidirectional authentication of ECUs.
In this section, we summarize the key protocol parameters and describe the procedures of session key generation as well as authentication message generation and verification. 

\vspace{0.1cm}
\noindent \textbf{Authentication protocol parameters.} 
In TACAN, both the ECU that transmits CAN messages with ID or priority equal to $k$ \footnote{In our earlier work \cite{ying2019tacan}, we assumed one message ID per ECU. In practice, however, an ECU may transmit CAN messages with different IDs. In this case, TACAN needs to be deployed for each message ID. } and the MN store a tuple $\langle MK^{(k)}, C_s^{(k)}, SK^{(k)}, C_m^{(k)} \rangle$, where
\begin{itemize}
    \item The master key $MK^{(k)}$ is a long term pre-shared key that is used to generate the session key; 
    
    \item The session key counter $C_s^{(k)}$ is a counter that is incremented by one at every vehicle start-up or when $C_m^{(k)}$ overflows; it is used to generate the session key;
    
    \item The session key $SK^{(k)}$ is a key used to generate authentication messages for CAN messages with ID=$k$;
    
    \item The authentication message counter $C_m^{(k)}$ is a counter that is incremented by one before being used to generate the next authentication message. 
\end{itemize}
We assume that $MK^{(k)}$, $SK^{(k)}$ and $C_{s}^{(k)}$ are securely stored in the TPMs of both the ECU and the MN.

    
    
    
    


\begin{figure}[h!]
    \centering
    \includegraphics[width=1\columnwidth]{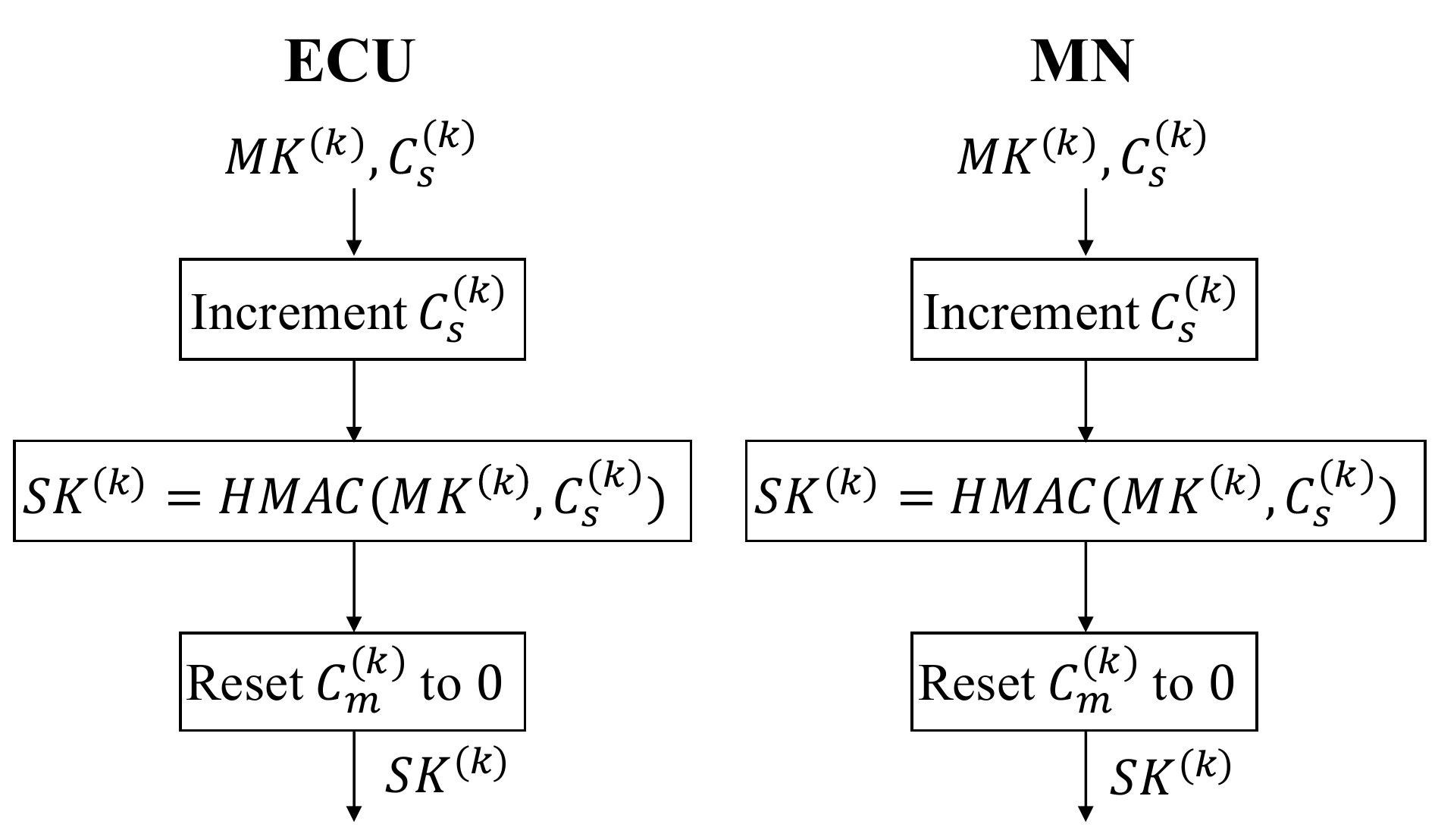}
    \caption{Session key generation for the ECU that transmits CAN messages with ID=$k$ and the MN. }
    \label{fig:session_key_gen}
\end{figure}

\vspace{0.1cm}
\noindent \textbf{Session key generation.}
The session key generation function takes a master key $MK^{(k)}$ and a session key counter $C_s^{(k)}$ as input and outputs a session key $SK^{(k)}$ (Fig.~\ref{fig:session_key_gen}).

As we can see, both the ECU and the MN first increment $C_s^{(k)}$ by one, i.e., $C_s^{(k)} \leftarrow C_s^{(k)}+1$. 
Then, they perform the following operation on $C_s^{(k)}$ using $MK^{(k)}$,
\begin{equation}
    SK^{(k)} = HMAC(MK^{(k)},C_s^{(k)}),\nonumber
\end{equation}
where $HMAC(\cdot)$ refers to the Hash-based Message Authentication Code (MAC) algorithm \cite{krawczyk1997hmac}.
Implementations of the protocol are free to use whichever hashing algorithm for HMAC and sizes of keys that are deemed strong enough.
One possible choice would be HMAC-SHA256 and 256 bits for keys. 
When a new session key is generated, $C_s^{(k)}$ is reset to zero. 

In the case of de-synchronized session key counters, a re-synchronization procedure such as the one in \cite{radu2016leia} will be needed, which may require message exchange between the ECU and the MN. 
The detailed design of such procedure is beyond the scope of this work. 

\begin{figure}[h!]
    \centering
    \includegraphics[width=.7\columnwidth]{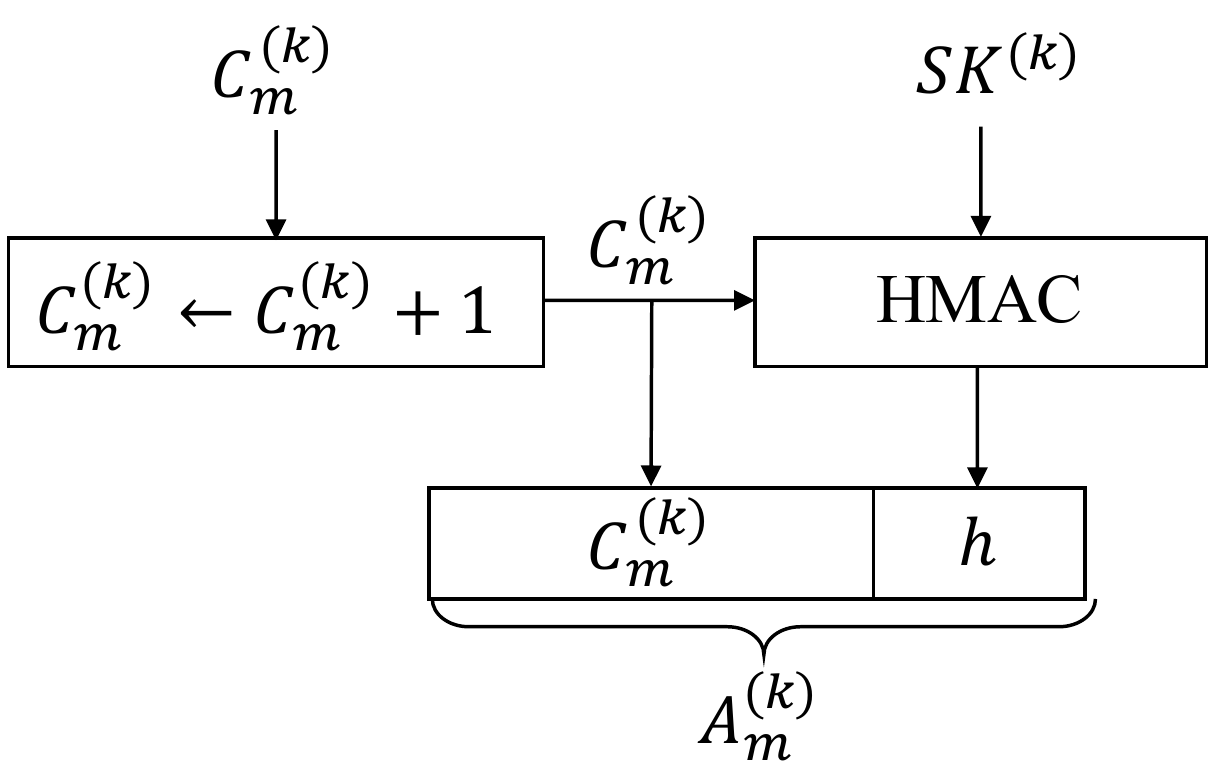}
    \caption{Illustration of authentication message generation. }
    \label{fig:auth_msg_gen}
\end{figure}

\vspace{0.1cm}
\noindent \textbf{Authentication message generation.}
As shown in Fig.~\ref{fig:auth_msg_gen}, the ECU that uses message ID $k$ increments $C_m^{(k)}$ by one and then generates the authentication message $A_m^{(k)}$ as follows: 
\begin{equation}
    A_{m}^{(k)} = C_m^{(k)} || HMAC(SK^{(k)}, C_m^{(k)}), \label{eq:A_m_structure} 
\end{equation}
where ``$||$'' denotes the concatenation of the counter value and the digest.
For the scope of this work, we assume that keys, counter values, and digests for TACAN are represented as bit strings. 

The generated authentication message is then transmitted using the covert channel and received by the MN.
Next, we will describe the verification procedure that happen on the MN side.

	


\vspace{0.1cm}
\noindent \textbf{Authentication message verification.} 
Algorithm~\ref{pseudo:auth_msg_ver} describes the authentication message verification procedure. 
For each received authentication message $\hat{A}_m^{(k)}$, the MN first extracts the counter value $\hat{C}_m^{(k)}$ and the corresponding digest $\hat{h}$ (Line~\ref{algo:ver:line1}). 
Then the MN increments its authentication message counter $C_m^{(k)}$ by one and compares it against $\hat{C}_m^{(k)}$ (Lines \ref{algo:ver:line2}-\ref{algo:ver:line3}). 
If the counter values are the same, the MN computes the expected digest $h$ (Line~\ref{algo:ver:compute_mac}) and compares it against $\hat{h}$ (Line~\ref{algo:ver:compare_mac}). 
If the digests are the same, the algorithm returns True, which indicates successful authentication message verification.
On the other hand, the mismatch of either the counter values or the digests means verification failure and indicates a possible CAN bus anomaly. 

\LinesNumbered
\begin{algorithm}[h!]
	\SetKwInOut{Input}{Input}
	\SetKwInOut{Output}{Output}
	\Input{$\hat{A}_m^{(k)}$, $SK^{(k)}$, $C_m^{(k)}$}
	\Output{True or False}
	
	$\hat{C}_m^{(k)}, \hat{h} \leftarrow \hat{A}_m^{(k)}$\; \label{algo:ver:line1}
	$C_m^{(k)} \leftarrow C_m^{(k)} + 1$\;\label{algo:ver:line2}
	\If{$\hat{C}_m^{(k)} = C_m^{(k)}$ \label{algo:ver:line3}}
	{
	    $h \leftarrow HMAC(SK^{(k)}, C_m^{(k)})$\;\label{algo:ver:compute_mac}
	    \If{ $\hat{h} = h$ \label{algo:ver:compare_mac}}
	    {
	       \textbf{return} True\;  
	    }\Else{
	        \textbf{return} False\;  
	    }
	}\Else 
	{
	    \textbf{return} False\; 
	}
\caption{Authentication message verification}
\label{pseudo:auth_msg_ver}
\end{algorithm}
\setlength{\textfloatsep}{3pt}


\vspace{0.1cm}
\noindent \textbf{Authentication frame structure.}
In TACAN, authentication messages are independent of CAN messages. 
They are encapsulated in authentication frames and transmitted through the covert channel.
Each authentication frame consists of four fields:
1) a SOF field, 2) a Data field, 3) a CRC field, and 4) EOF field. 
The SOF and EOF fields indicate the start and the end of an authentication frame, and the CRC field is used for error detection and ensures data integrity.

Inspired by the CAN frame structure (Fig.~\ref{fig:can_frame_structure}), we use a single bit 0 for the SOF and a bit string of 7 consecutive 1's for the EOF. 
We use an 8-bit CRC (CRC-8), which is also part of the AUTOSAR specifications \cite{autosar2017crc}, but more bits may be needed if necessary. 
Compared to the CAN frame, there are no arbitration, control, or ACK fields in the authentication frame, as they are no longer needed. 
When there are no authentication bits that need to be transmitted, bits of value one will be transmitted to fill the gap between frames. 

In order to avoid confusion due to possible appearances of the EOF in the frame, the bit stuffing technique will also be used, which inserts one bit of opposite polarity after five consecutive bits of the same polarity (including the preceding stuffed bit). 
The stuffed frame can then be destuffed by the receiver.
Note that the same bit stuffing technique is also used in the CAN protocol to maintain bit-level timing synchronization.


As shown in Eq.~(\ref{eq:A_m_structure}), each authentication message contains a counter value and its digest.
A 24-bit counter can already last for 46+ hours for 10-ms CAN messages before having an overflow even in per-message authentication. 
While we assume a 24-bit counter in this work, fewer bits may be used in practice due to two reasons.
First, authentication messages in TACAN are transmitted in a much lower frequency, which means that the counter overflow will take a much longer time.
Second, since it is a monotonic counter incremented by one for every frame, the transmitter may only transmit the last few bits of the counter value to keep the receiver synchronized in the case of occasionally corrupted frames. 

As for the digest, instead of transmitting the entire digest that are usually hundreds of bits long, the transmitter may truncate each digest to several bits to reduce the transmission time (e.g., using the least significant 8 bits or XORing all bytes together to create a condensed 8-bit version of the digest, as in~\cite{kurachi2014cacan,szilagy2008flexible}).
In practice, the number of bits in the shortened version should be chosen appropriately to achieve a desirable security level.

\subsection{IAT-Based Covert Channel}
\label{sec:IAT_based_covert_channel}
\begin{figure}[t!]
    \centering
    \includegraphics[width=1\columnwidth]{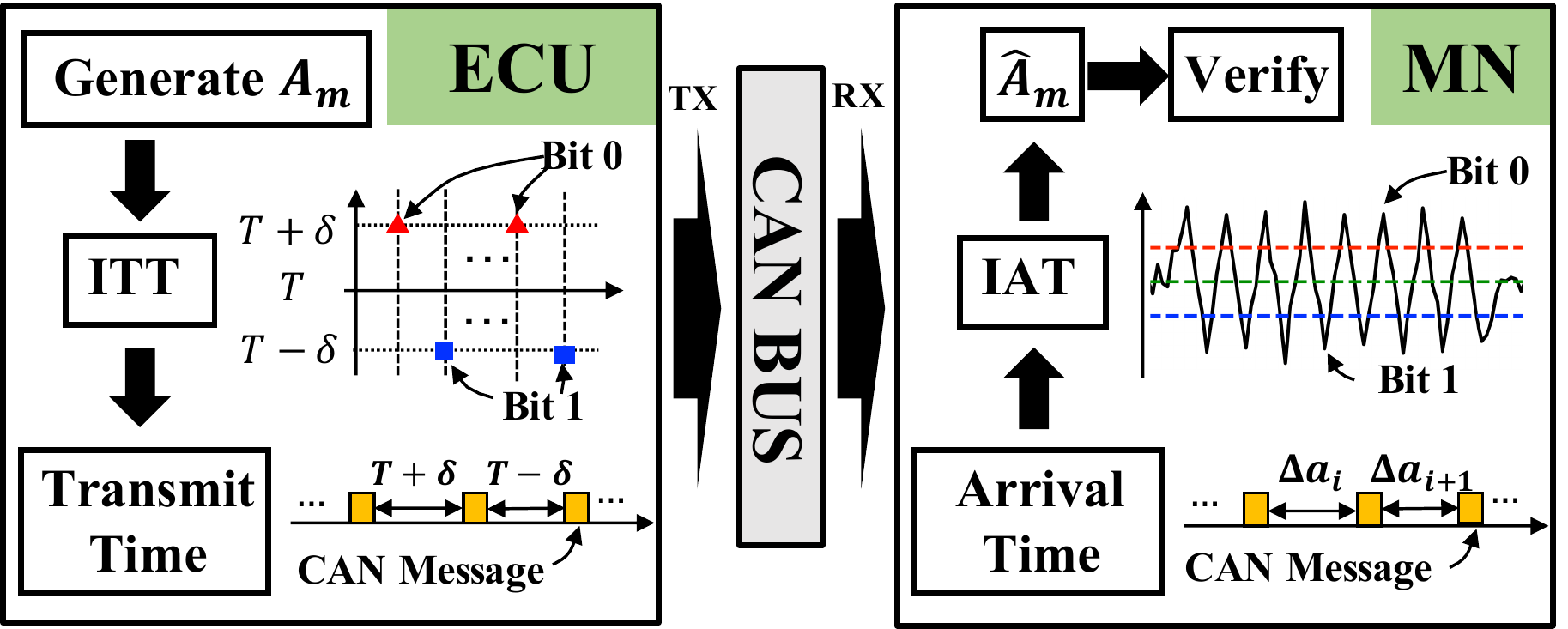}
    \caption{Illustration of the IAT-based covert channel. Each ECU embeds its authentication message $A_m$ into the ITTs of CAN messages, which can be observed and extracted by the MN through IATs. By verifying the received authentication message $\hat{A}_m$, the MN can verify the authenticity of the ECU. }
    \label{fig:timing_based_covert_channel}
\end{figure}

Fig.~\ref{fig:timing_based_covert_channel} illustrates the IAT-based covert channel for periodic CAN messages. 
In this covert channel, the ECU embeds the authentication bits into the ITTs of CAN messages, which can be extracted from the IATs by the MN. 
By verifying the received authentication message, the MN can authenticate the transmitter.

In the rest of this section, we will first motivate the design of the IAT-based covert channel through observations, and then present the modulation/demodulation schemes. 
After that, we will discuss the impact of two key parameters, that is, the window length $L$ and the added deviation $\delta$ on the bit error probability and the CAN bus schedulability.

\vspace{0.1cm}
\noindent \textbf{Observations.}
From Eq.~(\ref{eq:inter_arrival_time}), we observe that if an amount of deviation $\delta$ is added to (or subtracted from) the ITTs, the receiver will see a corresponding change in the IATs.
Hence, one simple scheme is to set $\Delta t_i=T+\delta$ for transmitting a bit $0$ and $\Delta t_i=T-\delta$ for transmitting a bit $1$.

\begin{figure}[t!]
    \centering
    \begin{subfigure}[b]{0.48\columnwidth}
        \includegraphics[trim=0cm 0.3cm 0cm 0cm,clip=true,width=\textwidth]{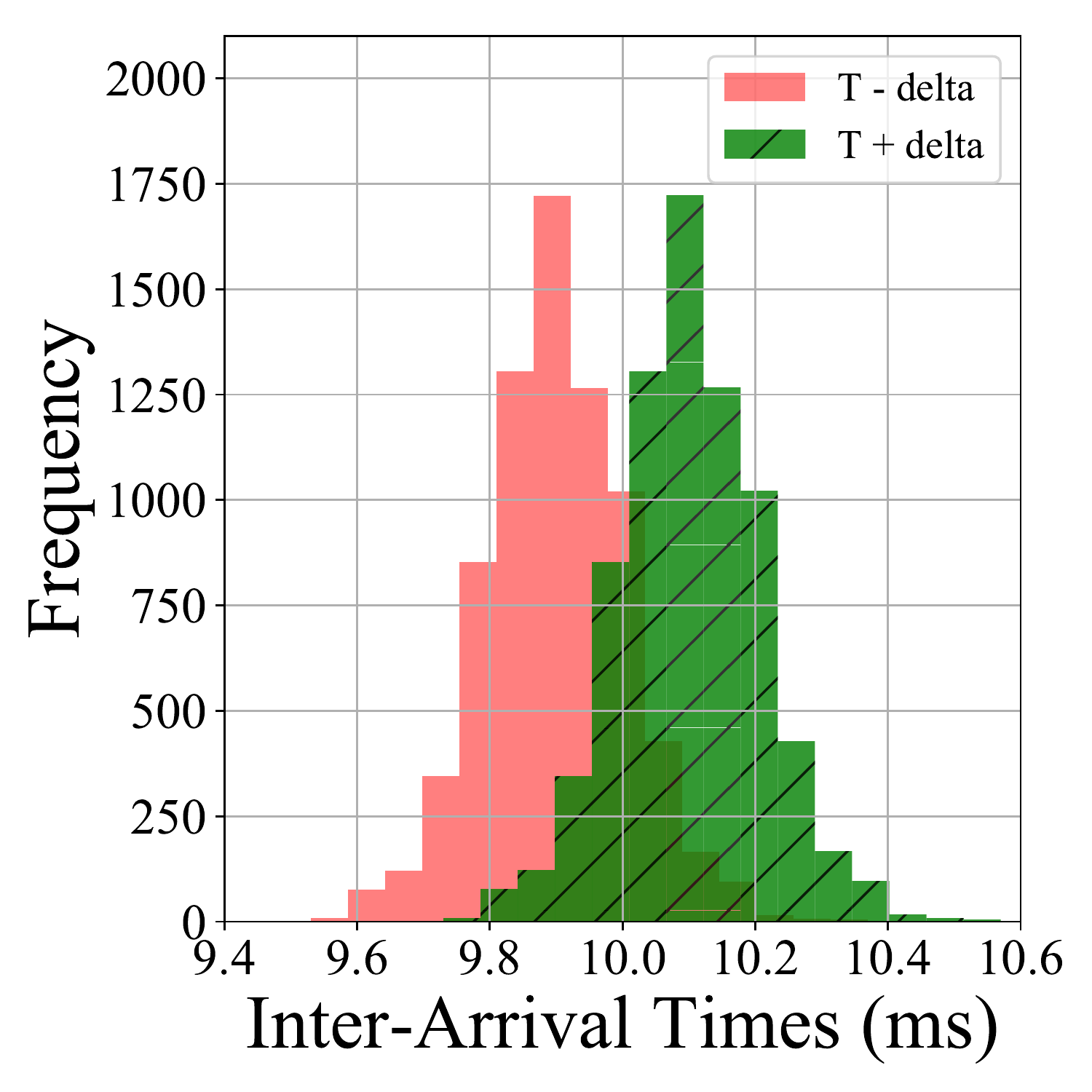}
        \caption{}
        \label{fig:example_iat_distribution_L-1}
    \end{subfigure}
    ~
    \begin{subfigure}[b]{0.48\columnwidth}
        \includegraphics[trim=0cm 0.3cm 0cm 0cm,clip=true,width=\textwidth]{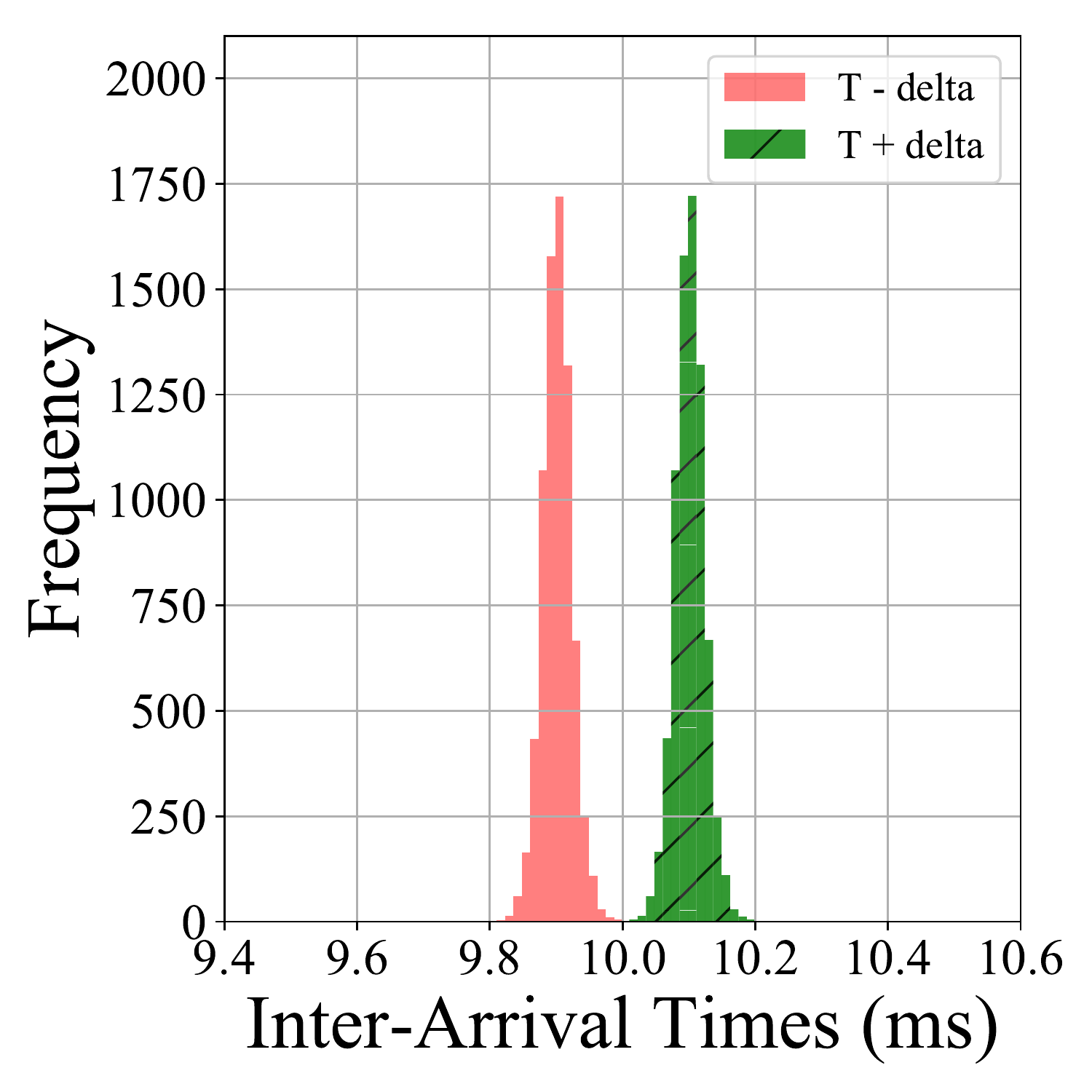}
        \caption{}
        \label{fig:example_iat_distribution_L-4}
    \end{subfigure}
    \caption{Example of IAT distributions of message 0x020 from the Toyota dataset at the receiver side (a) without running average ($L=1$) and (b) with running average ($L=4$).} 
    \label{fig:example_iat_distribution}
\end{figure}

Taking message 0x020 from the 2010 Toyota Camry \cite{toyota2010dataset} as an example, we plot its IAT distributions with added deviations in Fig.~\ref{fig:example_iat_distribution_L-1}, where $T=10$ ms and $\delta = 0.01T = 0.1$ ms. 
While adding a small deviation $\delta$ does lead to a shift in the IAT distribution, the two distributions are still overlapping with each other, indicating possible bit errors due to the noise caused by the CAN traffic. 

In order to reduce bit errors, one way is to average the IATs with a window length of $L$ at the receiver to smooth out the noise.
As we can see in Fig.~\ref{fig:example_iat_distribution_L-4}, 
performing running averages can effectively separate the two clusters from each other.
To support the computation of running averages at the receiver, the transmitter needs to transmit the same bit over $L$ consecutive ITTs. 
The above observations motivate the design of the IAT-based covert channel.

Note that when the probability of bit errors is reasonably small, one may also consider Error Correction Coding (ECC) techniques that are widely used in communications \cite{goldsmith2005wireless} to detect and recover occasional bit errors. 
Nevertheless, the use of ECC techniques will complicate the design of covert channels and introduce overheads.
As we will show later in Section~\ref{sec:evaluation}, since the probability of consecutive corrupted authentication frames is small, it is acceptable to not recover the bit errors in a particular frame.

\vspace{0.1cm}
\noindent \textbf{Modulation.} 
The transmitter uses the following modulation scheme to embed a bit into the ITTs of CAN messages,
\begin{equation*}
    \Delta t_j =
    \begin{cases}
        T + \delta, & \text{ if } b_i = 0\\
        T - \delta, & \text{ if } b_i = 1
    \end{cases},
\end{equation*}
where $j \in [iL, (i+1)L)$. 
This is essentially the Binary Phase Shift Keying (BPSK) modulation in the communication theory \cite{goldsmith2005wireless}.
While more than two levels may be used to increase the throughput, we consider only two levels in this work to minimize the impact on the CAN bus schedulability.

\vspace{0.1cm}
\noindent \textbf{Demodulation.} 
From Eq.~(\ref{eq:inter_arrival_time}), we know that the original IATs are $\Delta a_j = T/(1+S) + n_j$. 
When $\Delta t_j$ is changed to $T \pm \delta$, the receiver will observe
\begin{equation*}
    \Delta a_j =\frac{T + (2 b_i - 1) \delta}{1+S} + n_j,
\end{equation*}
where $b_i \in \{0, 1\}$ is the transmitted bit and $j \in [iL, (i+1)L)$.
In this work, we assume that $\eta_j$'s are i.i.d. Gaussian random variables, i.e., $\eta_j \sim N(d, \sigma_\eta^2)$.
Hence, we have $n_j=\eta_j-\eta_{j-1} ~\sim N(0, \sigma^2)$, where $\sigma^2=2\sigma_\eta^2$.

In order to demodulate the authentication message from the IATs, the MN needs to perform three steps: 1) computing running averages, 2) sampling, and 3) thresholding. 
In the first step, the MN computes the running averages of window length $L$ as below,
\begin{align}
    \Delta \bar{a}[i] &= \frac{1}{L} \sum_{j=0}^{L-1}\Delta a_{i+j} \nonumber \\
    &= \frac{1}{1+S} \left( \frac{1}{L}  \sum_{j=0}^{L-1}\Delta t_{i+j} \right) + \frac{1}{L}(\eta_{i+L-1} - \eta_{i-1}). \label{eq:running_average}
\end{align}
From Eq.~(\ref{eq:running_average}), we can see that the noise term is reduced by a factor of $L$ and the resulting variance of IATs is $2\sigma_{\eta}^2/L^2$.

After computing the running averages $\{\Delta \bar{a}[i]\}$, the MN needs to sample every $L$ values with a correct sampling offset $\tau^*$ and obtains $\{\Delta \bar{a}[iL + \tau^*]\}$.
From the communication theory \cite{goldsmith2005wireless},
we know that the optimal receiver minimizes the bit error probability $P_e$ by selecting the output $\hat{b}[i]$ that maximizes $1-P_e=\text{Pr}(b_i \text{ sent } | \Delta \bar{a}[iL + \tau^*] \text{ received})$.
If we assume that bits 0 and 1 are transmitted with equal probabilities, then the maximum likelihood decision criterion is
\begin{equation*}
    \hat{b}[i] =
    \begin{cases}
        0, & \text{ if } \Delta \bar{a}[iL+\tau^*] \geq \Gamma,\\
        1,  & \text{ otherwise },
    \end{cases}
\end{equation*}
where $\Gamma=  \frac{T}{1+S}$ (Eq.~(\ref{eq:IAT_mean})), that is, the decision threshold is the mean of original IATs or the observed message period. 
Lastly, the MN can decode $A_m$ from the bit string $\{\hat{b}[i]\}$. 

Note that it is not necessary for the receiver to be synchronized with the transmitter for the sampling purposes, because the receiver can determine the correct sampling offset $\tau^*$ by itself as follows, 
\begin{equation}
    \tau^* = \arg \max_{\tau} \sum_i \left|\Delta \bar{a}[iL+\tau] - \Gamma \right|,
    \label{eq:optimal_tau}
\end{equation}
where $\tau\in \{0,1,..,L-1\}$.
That is, $\tau^*$ is the integer value that maximizes the total distance between each sample and the decision threshold. 

\vspace{0.1cm}
\noindent \textbf{Impact of $\delta$ and $L$ on bit error probability.}
From our previous discussion, we see that $L$ and $\delta$ are the two key parameters that affect the bit error performance of the IAT-based covert channel. 
In this section, we analytically model the bit error probability $P_e$ as a function of $L$ and $\delta$. 

First of all, based on our previous assumption of Gaussian noise, we have 
\begin{equation}
    \Delta \bar{a}[iL + \tau^*] ~\sim N\left( \frac{T + (2b_i-1)\delta }{1+S}, \frac{\sigma^2}{L^2} \right), \label{eq:distribution_of_sampled_IAT}
\end{equation}
where $b_i \in \{0,1\}$ and $\sigma^2 = 2\sigma^2_\eta$ (Eq.~(\ref{eq:noise_variance})).
Let $P_{e,0}$ and $P_{e,1}$ be the bit error probabilities conditioned on bits 0 and 1 being transmitted, respectively. 
Then we have
\begin{align*}
P_{e,0} &= \text{Pr}(\Delta \bar{a}[iL+\tau^*] < \Gamma | b_i = 0), \\
P_{e,1} &= \text{Pr}(\Delta \bar{a}[iL+\tau^*] \geq \Gamma | b_i = 1).
\end{align*}
According to Eq.~(\ref{eq:distribution_of_sampled_IAT}), we know that
\begin{align}
    P_{e,0} = \text{Pr}\left(Z < \frac{ -\delta/(1+S)}{\sigma/L}\right) = Q \left( \frac{L \delta}{(1+S)\sigma } \right) \nonumber,
\end{align}
where $Z\sim N(0,1)$ is the standard Gaussian random variable and $Q(x)\triangleq \frac{1}{\sqrt{2\pi}} \int_{z}^{\infty} exp(-\frac{z^2}{2}) dz$ is the Q function. 
Due to the symmetry, we have $P_{e,0}=P_{e,1}$. 

Assuming equally likely probabilities of bits 0 and 1, i.e., $P_0=P_1=0.5$, the total error bit probability is
\begin{equation}
    P_e = P_{e,0}\cdot P_0 + P_{e,1} \cdot P_1 =Q \left( \frac{L \delta}{(1+S)\sigma} \right). \label{eq:bit_error_probability}
\end{equation}

From Eq.~(\ref{eq:bit_error_probability}), we see that increasing $L$ and $\delta$ would have the same effect on $P_e$. 
Nevertheless, increasing $L$ will reduce the covert channel throughput, whereas increasing $\delta$ could potentially affect the schedulability of the CAN bus, as we will explain in the next section.
Therefore, in the case of fixed $\delta$, it makes full sense to choose the smallest $L$ value, while keeping $P_e$ less than or equal to a given limit $\epsilon$. 
From Eq.~(\ref{eq:bit_error_probability}), we have
\begin{equation}
    L_{\min} = \left\lceil \frac{(1+S)\sigma}{\delta} Q^{-1}(\epsilon) \right\rceil,
\end{equation}
where $\lceil \cdot \rceil$ is the ceiling function.

\vspace{0.1cm}
\noindent \textbf{Impact of $\delta$ on CAN bus schedulability:}
Timing on the CAN bus is closely related to its schedulability.
A CAN bus is schedulable if and only if all the messages on the bus are schedulable, and a message with ID $k$ is schedulable if and only if its worst-case response time (denoted as $R_k$) is less than or equal to its deadline (denoted as $D_k$). 

According to \cite{davis2007controller}, $R_k$ is defined as the longest time from the initiating event (that puts the message in the transmission queue) occurring to the message being received by the nodes that require it.
It consists of three parts: 1) the queuing jitter, 2) the queuing delay, and 3) the transmission time. 
The queuing delay further consists of the blocking delay (due to ongoing transmissions of lower priority messages) and the interference (due to the arbitration process when competing with higher priority messages).

Since the IAT-based covert channel introduces a deviation of $\delta$, its effect is equivalent to decreasing the message period (defined as the minimum inter-transmission time) and increasing the queuing jitter. 
By applying the schedulability analysis in \cite{davis2007controller}, we can show that the impact of $\delta$ on the worst-case response time
is threefold: 1) increase in the queuing jitter by a fixed amount ($\delta$), 2) increase in the blocking delay by a bounded amount of time, and 3) increase the message transmission time of higher priority messages by a certain percentage ($\delta/T$). 
Therefore, to achieve the effective use of the IAT-based covert channel, TACAN parameters (notably $\delta$) need to be experimentally obtained and fine tuned prior to deployment to ensure the schedulability of the CAN bus.
A detailed discussion is provided in Appendix A. 

\subsection{LSB-Based Covert Channel}
\label{sec:lsb_based}
\begin{figure}[t]
    \centering
    \includegraphics[width=1\columnwidth]{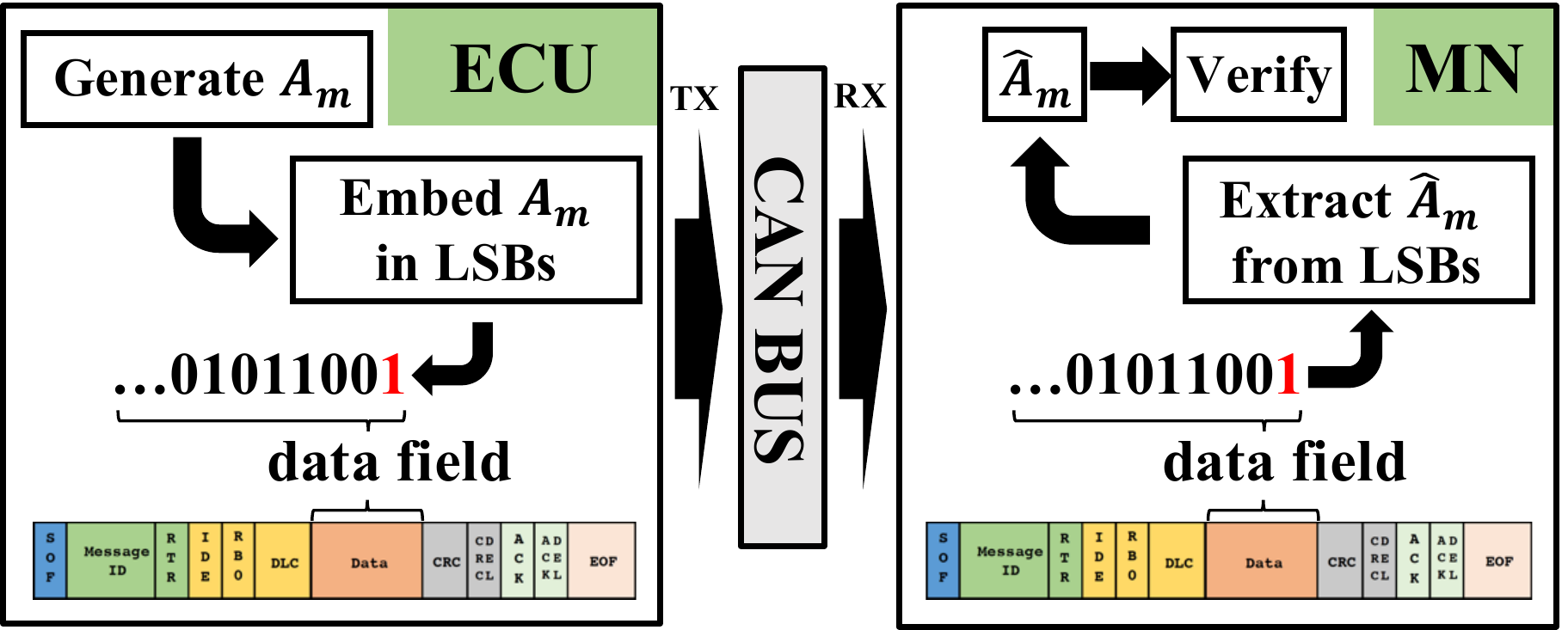}
    \caption{Illustration of LSB-based covert channel. The transmitting ECU embeds the authentication message into the LSBs of multiple normal CAN messages (with the same ID), which can be extracted and verified by the Monitor Node.}
    \label{fig:lsb_scheme}
\end{figure}

In this section, we present the LSB-based covert channel, which 
embeds the authentication messages inside the LSBs of the data payload of normal CAN messages transmitted by an ECU, as illustrated in Fig.~\ref{fig:lsb_scheme}.
Unlike the IAT-based covert channel, the LSB-based covert channel is also applicable to aperiodic CAN messages.
For the scope of this work, we use the CAN data frames to develop our methodology.

\vspace{0.1cm}
\noindent \textbf{Observations.}
In order to transmit authentication messages over the CAN bus, it is common to leverage the existing fields of a CAN message, such as the data field (at least one byte) and the extended ID field, or simply introduce additional CAN messages~\cite{kurachi2014cacan,hazem2012lcap,groza2012libra}.
In practice, however, there may not be any unused bytes in the CAN message or the CAN bus is already heavily loaded, which makes the above approaches difficult to deploy. 
In TACAN, the objective is to authenticate the transmitter instead of each CAN message, and thus authentication messages are transmitted much less frequently. 
If we can spread the bits of an authentication message across multiple CAN messages, each of which carries only a few authentication bits (e.g., one or two bits), we can then alleviate the payload shortage and avoid traffic overheads.

On the other hand, we also observe that CAN messages are often used to convey sensor values, most of which are floating values represented in bits. 
Therefore, the $L$ LSBs (e.g., $L=1$ or $2$) of each such CAN message may be used for transmitting authentication bits without causing significant degradation in accuracy. 

Taking the 2010 Toyota Camry \cite{toyota2010dataset} as an example, there are at least $7$ messages out of $42$ that carry sensor values (e.g., wheel speeds, engine speed, vehicle speed, odometer, brake pressure, steering angle) \cite{toyota2010dataset}. 
We would expect more CAN messages that carry sensor values in newer automobiles.
The above observations motivate our design of LSB-based covert channels.

\vspace{0.1cm}
\noindent \textbf{Embedding $A_m$ to LSBs.} 
The embedding procedure is considered as a sub-layer between the Application and the Data Link layers. 
The basic idea is as follows: for each message $A_m$, the transmitter substitutes the least significant $L$ bits of the CAN message with the next $L$ bits in $A_m$. 
No modification is needed if the $L$ bits happen to be the same.

As provided in Algorithm \ref{pseudo:ecu_tx}, an authentication frame $A_f$ is first constructed from $A_m$  (Line~\ref{algo:lsb:A_f}). 
When the new data field content $d_{field}$ is received from the upper (Application) layer, the least significant $L$ bits of the selected byte $\beta$ in $d_{field}$, denoted as $bin(d_{field}(\beta))_L$, are compared with the next $L$ bits of $A_f$: if the bits of interest are different, then the substitution occurs; otherwise, $d_{field}$ is not modified (Lines~\ref{algo:lsb:line3}-\ref{algo:lsb:line7}).
The same process is repeated for each new authentication message. 


\LinesNumbered
\begin{algorithm}[h!]
	\SetKwInOut{Input}{Input}
	\SetKwInOut{Output}{Output}
	Construct $A_f$ from $A_m$ (with bit stuffing)\;
	\label{algo:lsb:A_f}
	$i \leftarrow 0$\; 
	\While{$i < length(A_f)$ \label{algo:lsb:line3}} 
	{
	    Receive $d_{field}$ from the upper (Application) layer\;
	    \If{$bin(d_{field}(\beta))_L \neq A_f[i:i+L-1]$}
		{
		    $bin(d_{field}(\beta))_L = A_f[i:i+L-1]$\;
		}
        $d_{frame} \leftarrow d_{field}(\gamma)$\;
		$i \leftarrow i + L$\; \label{algo:lsb:line7}
		Send $d_{field}$ to the lower (Data Link) layer\;
	}

	\textbf{return}\;
\caption{Embedding $A_m$ to LSBs}
\label{pseudo:ecu_tx}
\end{algorithm}
\setlength{\textfloatsep}{3pt}

\vspace{0.1cm}
\noindent \textbf{Extracting $A_m$ from LSBs.} 
On the receiver side, the MN extracts the $L$ LSBs from every received CAN message and reconstructs the authentication message.
If the MN fails to verify the authentication message, it will raise an alert that indicates possible compromise of the transmitting ECU or malicious exploitation of the CAN bus.

The extracting procedure is described in Algorithm \ref{pseudo:mon_rx}. 
Once the MN detects the SOF flag, it will start extracting the $L$ LSBs from the designated byte $d_{field}(\beta)$ and appending them to $\hat{A}_f$ (Lines~\ref{algo:lsb:extract:line1}-\ref{algo:lsb:extract:line5}). 
When the EOF flag is detected in $\hat{A}_f$, the MN will perform bit destuffing on $\hat{A}_f$ and extract the authentication message $\hat{A}_m$ (Line~\ref{algo:lsb:extract:line6}). 
Then it will listen for the SOF of the next incoming frame. 

\LinesNumbered
\begin{algorithm}[h!]
	\SetKwInOut{Input}{Input}
	
	\If{SOF is detected\label{algo:lsb:extract:line1}} 
	{
	    Initialize $\hat{A}_f$ to an empty bit string\;
	    \While{EOF is not detected in $\hat{A}_f$}{
	        Receive $d_{field}$ from the Data Layer layer\;
	        Append $bin(d_{field}(\beta))_L$ to $\hat{A}_f$\; \label{algo:lsb:extract:line5}
	    }
	    Destuff $\hat{A}_f$ and extract $\hat{A}_m$\;  \label{algo:lsb:extract:line6}
	}
	
	\textbf{return} $\hat{A}_m$\;
	
	\caption{Extracting $A_m$ from LSBs}
	\label{pseudo:mon_rx}
\end{algorithm}
\setlength{\textfloatsep}{5pt}
\subsection{Hybrid Covert Channel}
\label{sec:hybrid}
Since the IAT-based and LSB-based covert channels are orthogonal to each other, it leads to a natural question whether the two covert channels can be combined to construct a hybrid channel to achieve a larger throughput. 
In this hybrid covert channel, the transmitter will split the authentication message into two parts and transmit them through the two covert channels separately. 
Then the receiver will receive and reassemble the two parts into one piece.
In this work, we are interested in answering the following question: how to choose the splitting ratio $\alpha$ ($0\leq\alpha\leq 1$) such that the two parts will be transmitted roughly over the same duration. 

Given $N_m$ bits for the authentication message and $N_o$ bits for other fields (SOF, CRC, and EOF), if we ignore bit stuffing, then there are a total of $\lceil\alpha N_m \rceil + N_o$ bits and $\lceil(1-\alpha) N_m \rceil + N_o$ bits transmitted over the IAT-based and LSB-based covert channels, respectively.

If the IAT-based covert channel is using a window length of $L_{1}$ and the LSB-based covert channel is using $L_2$ LSBs, then requiring the same transmission duration means
\begin{equation*}
    (\lceil\alpha N_m \rceil + N_o)L_1 = (\lceil(1-\alpha) N_m \rceil + N_o) /L_2.
\end{equation*}
Ignoring the ceiling function, we have \begin{equation}
    \alpha = \frac{N_m + N_o(1 - L_1 L_2)}{N_m(1+L_1 L_2)} \times 100\%. \label{eq:splitting_ratio}
\end{equation}
While $\alpha$ computed in in Eq.~(\ref{eq:splitting_ratio}) is not the optimal value $\alpha^*$ that leads to two partitions of equal length after bit stuffing, it will be close to $\alpha^*$, which provides a reasonable starting point for further iteration.
Note that $\alpha$ is less than or equal to $0$, it means that only the LSB-based covert channel should be used.

As an example, suppose $N_m=32$, $N_o=16$, $L_1=1$ and $L_2=1$. 
In this case, the two covert channels have similar throughput and we have $\alpha = 50.0\%$.
Nevertheless, if $L_1$ is increased from $1$ to $2$,  $\alpha$ quickly drops to $16.7\%$.
With $L_1=3$, $\alpha$ becomes $0$, which means that it is not advantageous to deploy the hybrid covert channel.

\section{Evaluation}
\label{sec:evaluation}
In this section, we conduct extensive experiments to evaluate the performance of the proposed covert channels and TACAN using real-world datasets. 
We first describe our real vehicle CAN bus testbed based on the UW EcoCAR (a Chevrolet Camaro 2016) and demonstrate the proposed IAT-based and LSB-based covert channels (Section~\ref{sec:eval:testbed_validation}). 
We then evaluate the bit error performance (Section~\ref{sec:eval:bit_error}), throughput (Section~\ref{sec:eval:throughput}), detection performance (Section~\ref{sec:eval:detection}), and accuracy loss (Section~\ref{sec:eval:accuracy_loss}) of TACAN using the EcoCAR datatset as well as the publicly available Toyota dataset collected from the Toyota Camry 2010 \cite{toyota2010dataset}. 

\subsection{Testbed Validation} 
\label{sec:eval:testbed_validation}
\begin{figure}[t!]
\centering
\includegraphics[trim=0cm 0.4cm 0cm -0.5cm,clip=true,width=1\columnwidth]{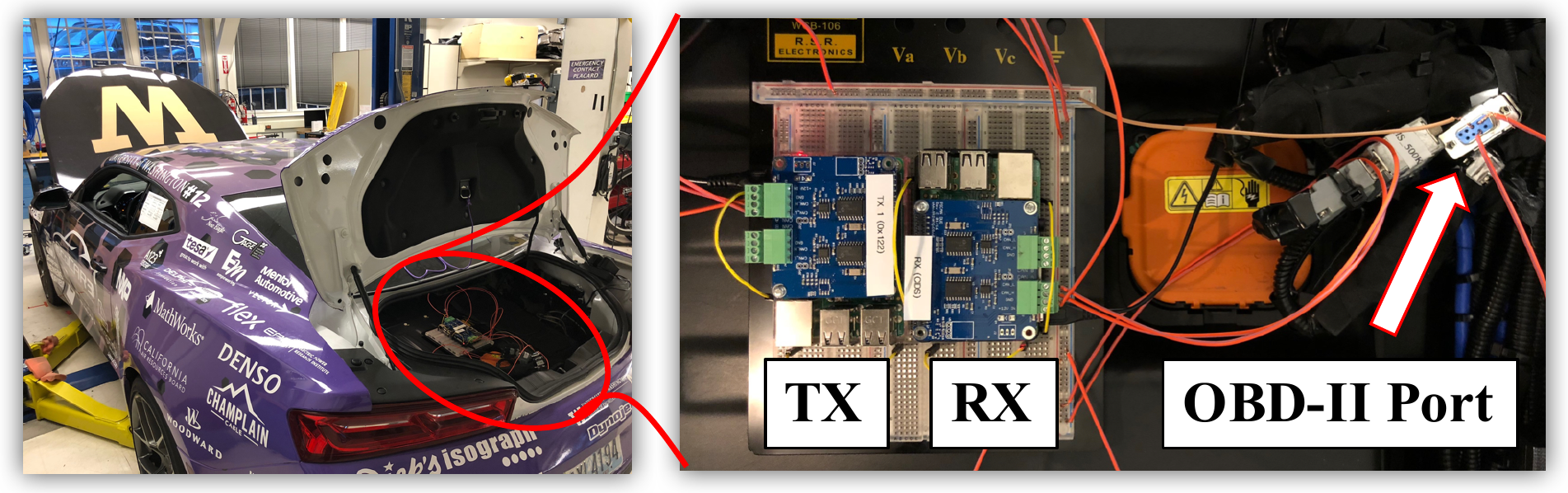}
\caption{Illustration of  EcoCAR testbed. Raspberry Pi-based testbed ECUs are connected to the OBD-II port at the back of the EcoCAR.
}
\label{fig:testbed_setup}
\end{figure}

Fig.~\ref{fig:testbed_setup} illustrates our EcoCAR testbed that consists of the UW EcoCAR and testbed ECUs, which are connected via the On-Board Diagnostics (OBD-II) port
The EcoCAR hosts 8 stock ECUs and two experimental ECUs.
A total of 2500+ messages with 89 different IDs are exchanged on the CAN bus every second.
All messages are periodic with periods ranging from 10 ms to 5 sec. 

Each testbed ECU consists of a Raspberry Pi 3 and a PiCAN 2 board (using a MCP2515 CAN controller and a MCP2551 CAN transceiver). 
The Raspberry Pi-based ECU is programmed to be a receive-only device that records CAN messages using SocketCAN \cite{socketCAN}.
During data collection, the EcoCAR is in the park mode in an isolated and controlled environment for safety purposes, but all in-vehicle ECUs are functional and actively exchange CAN messages. 

To demonstrate the proposed IAT-based and LSB-based covert channels, we use the testbed ECU to record the timestamps of 100-ms message 0x22A from the EcoCAR testbed and replay them in our experiments. 
In the ideal case, the TACAN transmitter will add $\pm \delta$ to the ITTs, which correspond to changes of $\pm \delta/(1+S)$ at the receiver side. 
Since both $\delta$ and $S$ are very small, we approximate $\delta/(1+S)$ as $\delta$ and program the TACAN transmitter to directly modify the recorded IATs.

\begin{figure}[t!]
    \centering
    \begin{subfigure}[b]{1\columnwidth}
        \includegraphics[trim=0cm 0.4cm 0cm 0cm,clip=true,width=1\textwidth]{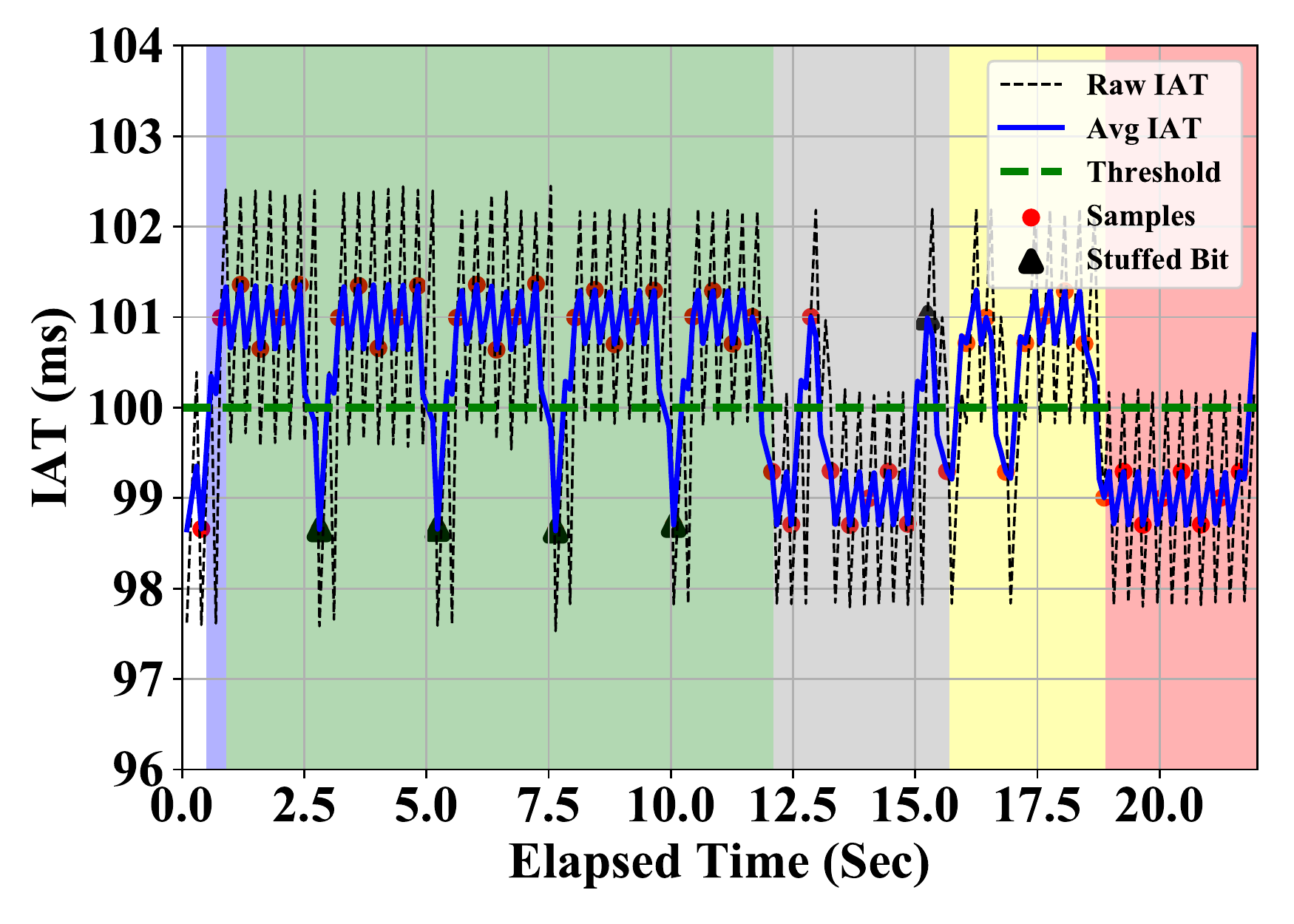}
        \caption{IAT-based covert channel}
        \label{fig:example_of_iat_based_covert_channel_0x22a_ecocar}
    \end{subfigure}
    ~
    \begin{subfigure}[b]{1\columnwidth}
        \includegraphics[trim=-0.8cm 0.4cm 0.4cm 0cm,clip=true,width=.98\textwidth]{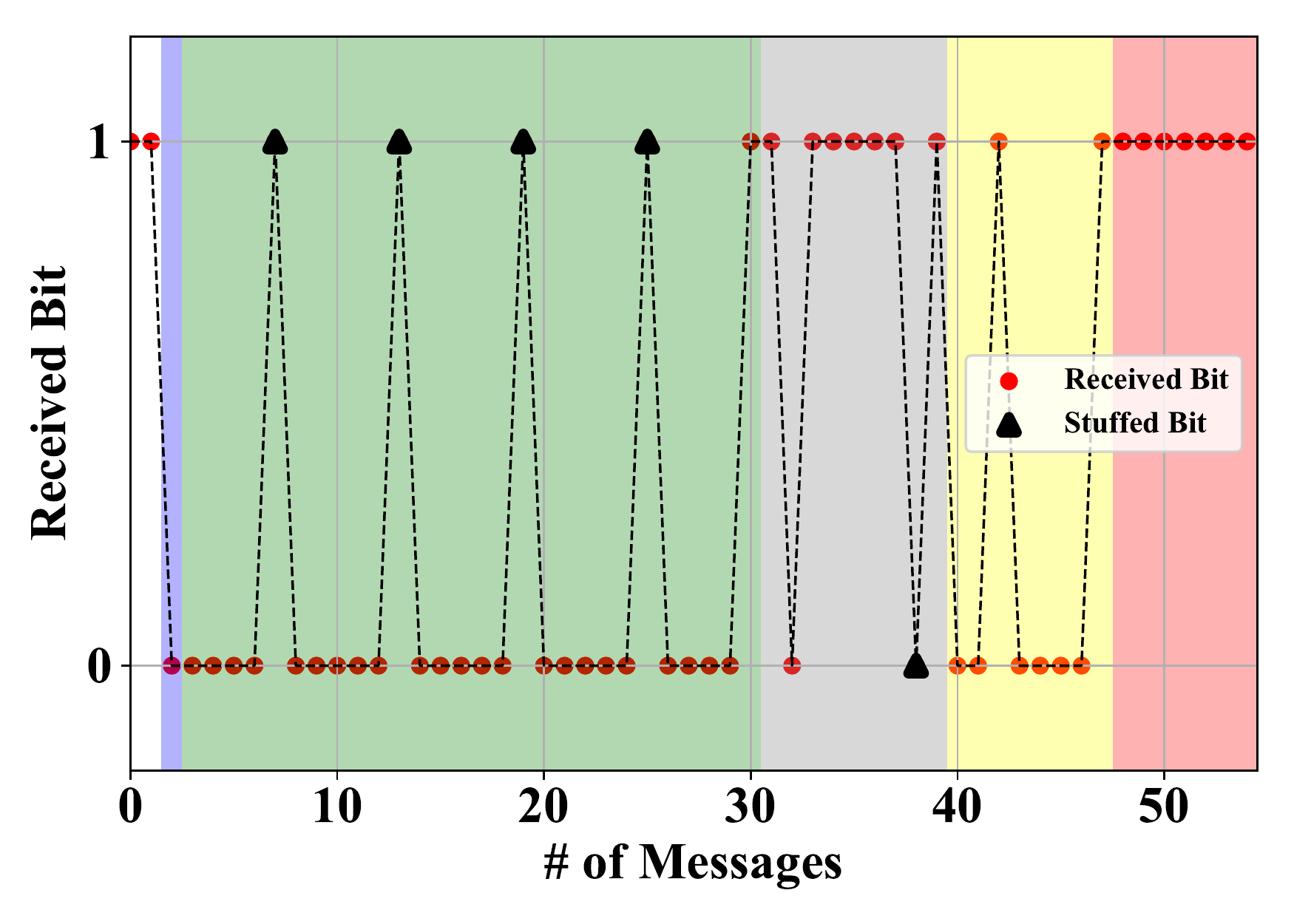}
        \caption{LSB-based covert channel}
        \label{fig:example_of_lsb_based_covert_channel_0x22a_ecocar}
    \end{subfigure}
    \caption{Example of received authentication frame for message 0x22A from the EcoCAR testbed in (a) the IAT-based covert channel and (b) the LSB-based covert channel. The fields for SOF, counter and digest values (Data), CRC, and EOF are colored in blue, green, gray, yellow, and red, respectively. The stuffed bits are marked as black triangles. 
    } 
    \label{fig:example_of_covert_channel_0x22a}
\end{figure}

Fig.~\ref{fig:example_of_iat_based_covert_channel_0x22a_ecocar} provides an example of the received authentication frame in the IAT-based covert channel.
In this example, the 24-bit counter has a value of 1, $\delta = 0.01T$, and $L=4$. 
We observe that without the running average, the received IATs can be very noisy, which leads to possible bit errors, whereas computing running averages can effectively smooth out the noise. 
We also observe that the actual length of the transmitted frame can exceed the original length due to the inserted stuffed bits.

Fig.~\ref{fig:example_of_lsb_based_covert_channel_0x22a_ecocar} provides an example of the received authentication frame in the LSB-based covert channel when a single LSB is used ($L=1$).
Compared to the IAT-based covert channel in Fig.~\ref{fig:example_of_iat_based_covert_channel_0x22a_ecocar}, the LSB-based covert channel is noise-free, and each authentication frame can be transmitted in a shorter duration. 

\subsection{Bit Error Performance}
\label{sec:eval:bit_error}
In this section, we discuss the bit error performance of the IAT-based and LSB-based covert channels. 
In addition to the EcoCAR dataset, we also use the Camry Toyota 2010 dataset \cite{toyota2010dataset}.
It consists of 43 distinct messages, and the majority of them are periodic with periods ranging from 10 ms to 5 sec.

\vspace{0.1cm}
\noindent \textbf{IAT-based covert channel.}
In this experiment, we select a subset of 9 representative messages from the Toyota dataset and the EcoCAR testbed with different message ID levels, periods, and noise levels, as listed in Table~\ref{table:representative_messages}. 
Our objective is to compare the analytical bit error probabilities ($P_e$) in Eq.~(\ref{eq:bit_error_probability}) against the experimental values and study the impact of $L$. 
We set $\delta =0.01T$ to minimize the impact on the CAN bus.  
Since $P_e$ is a function of the standard deviation of IATs $\sigma$ (Eq.~(\ref{eq:bit_error_probability})), we only consider IATs within $\pm 20\%$ of the average when computing $\sigma$ to minimize the adverse impact of outliers. 
To obtain the experimental $P_e$, we add $\pm \delta$ to the IATs and compute the running averages of window length $L$. 
We then compute the percentage of the averaged IATs that lead to bit errors as the experimental $P_e$.

\begin{table}[t!]
    \footnotesize
	\centering
	\caption{Statistics of IATs of representative messages from the Toyota dataset and the EcoCAR testbed.}
	\label{table:representative_messages}
    \begin{tabular}{|c|c|c|c|}
    \hline
    \multirow{2}{*}{Msg ID} & \multirow{2}{*}{\begin{tabular}[c]{@{}c@{}}Period\\ (sec)\end{tabular}} & \multirow{2}{*}{\begin{tabular}[c]{@{}c@{}}Standard deviation\\ (Norm. by period)\end{tabular}} & \multirow{2}{*}{Source} \\
     &  &  &   \\ \hline
    0x020 & 0.01 & 1.1\% & Toyota \\ \hline
    0x0B4 & 0.02 & 0.5\% & Toyota \\ \hline 
    0x224 & 0.03 & 0.9\% & Toyota \\ \hline
    0x620 & 0.3 & 1.0\% & Toyota \\ \hline
    0x0D1 & 0.01 & 2.7\% & EcoCAR \\ \hline
    0x185 & 0.02 & 1.6\% & EcoCAR \\ \hline
    0x22A & 0.1 & 1.2\% & EcoCAR \\ \hline
    0x3FB & 0.25 & 1.4\% & EcoCAR \\ \hline
    0x4D1 & 0.5 & 1.4\% & EcoCAR \\ \hline
    \end{tabular}
    \normalsize
\end{table}

\begin{figure}[t!]
    \centering
    \begin{subfigure}[b]{1\columnwidth}
        \includegraphics[trim=0cm 0.5cm 0cm 0cm,clip=true,width=1\textwidth]{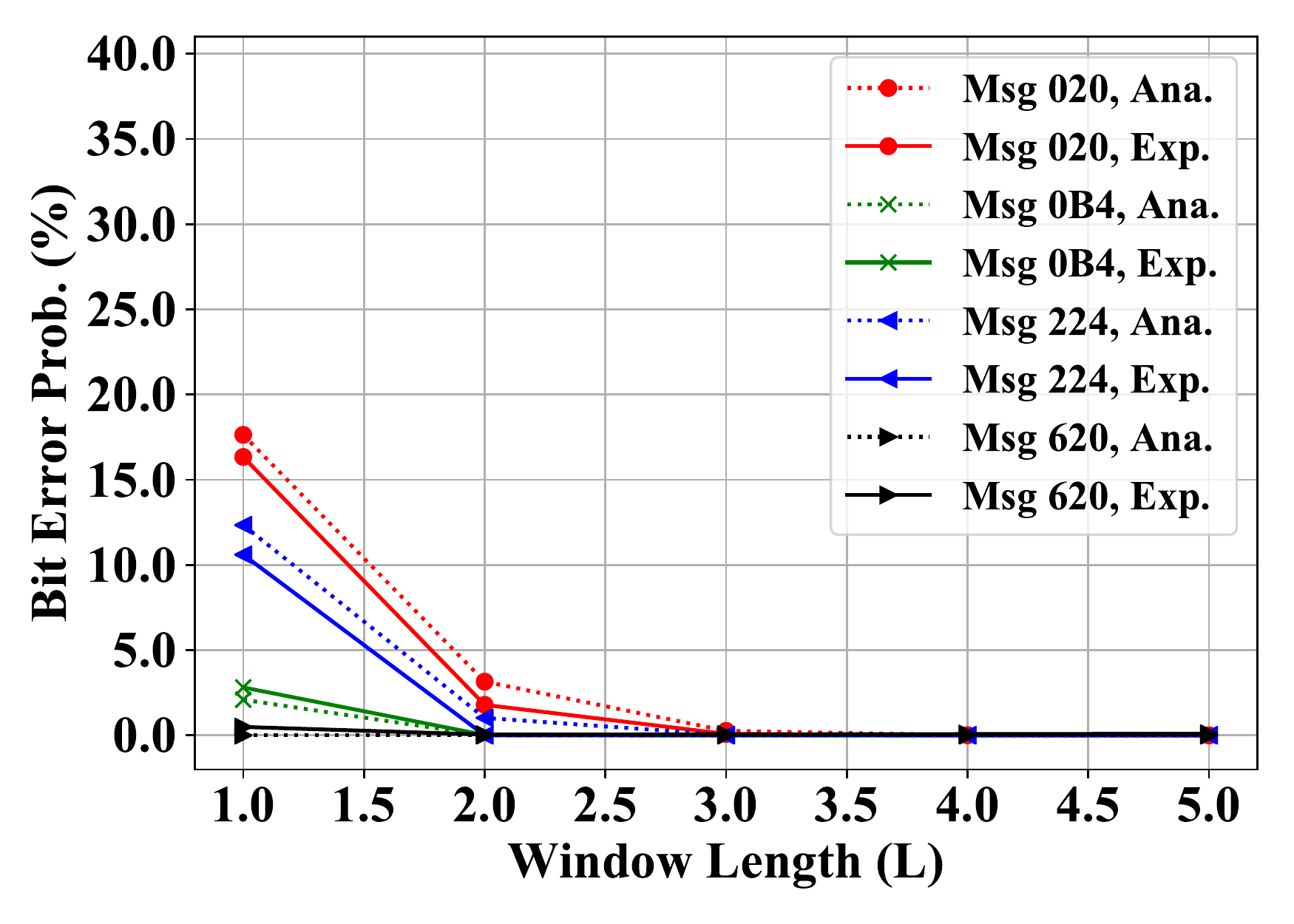}
        \caption{Toyota dataset}
        \label{fig:bit_error_prob_toyota}
    \end{subfigure}
    ~
    \begin{subfigure}[b]{1\columnwidth}
        \includegraphics[trim=0cm 0.5cm 0cm 0cm,clip=true,width=1\textwidth]{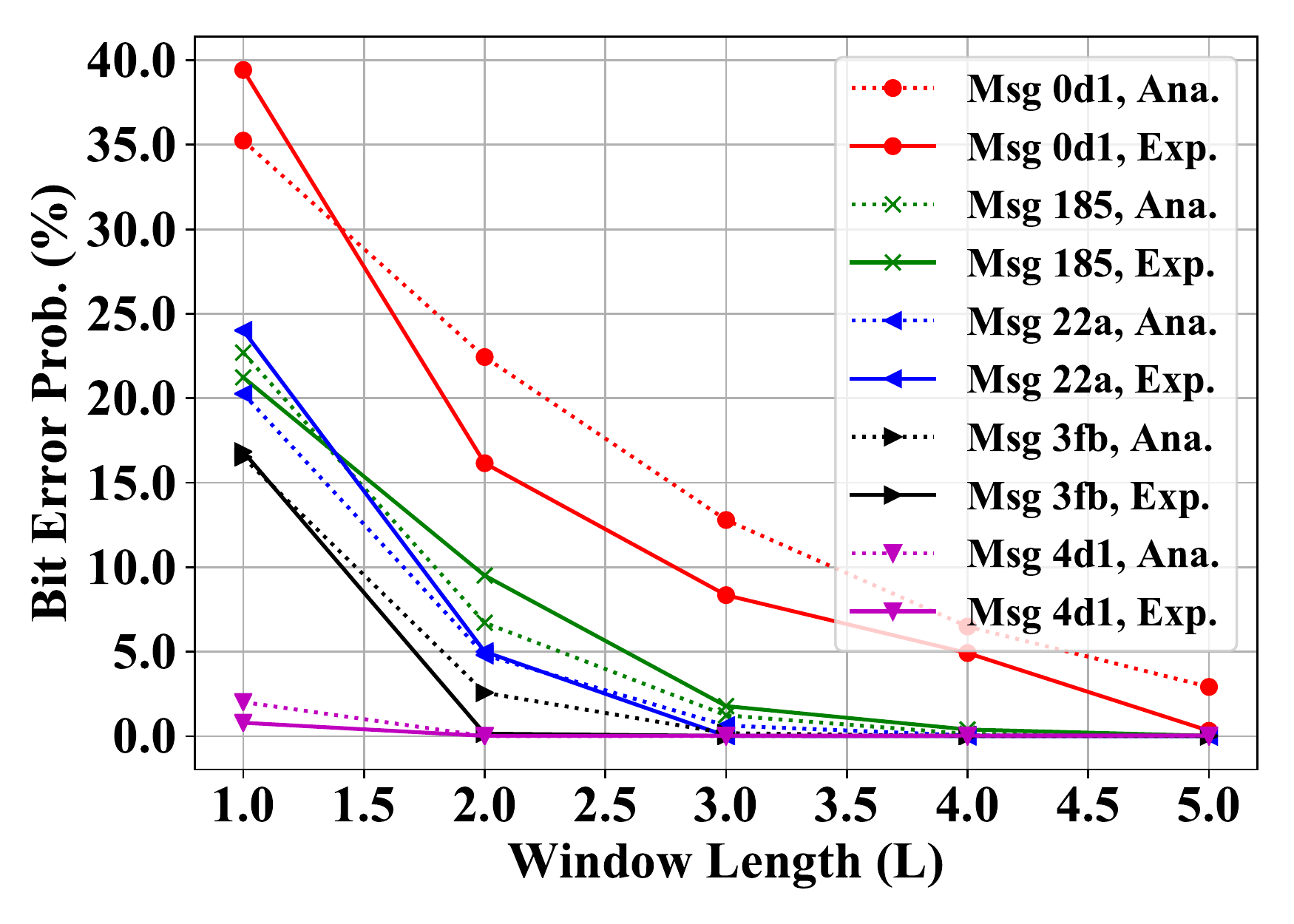}
        \caption{EcoCAR testbed}
        \label{fig:bit_error_prob_ecocar}
    \end{subfigure}
    \caption{Comparison of analytical and experimental bit error probabilities for selected messages from (a) the Toyota dataset and (b) the EcoCAR testbed. We observe that increasing $L$ will reduce the bit error probability. We also obesrve that our analysis provides a good estimate of the experimental bit error probability.
    } 
    \label{fig:bit_error_prob}
\end{figure}

As shown in Fig.~\ref{fig:bit_error_prob}, the bit error performance of the IAT-based covert channel varies a lot among different messages. 
In general, $P_e$ is very high with $L= 1$, but it quickly drops to as $L$ increases for all messages, which  demonstrates the effectiveness of the running average. 
For the four Toyota messages, the experimental $P_e$ is less than $0.1\%$ with $L=3$, and the experimental $P_e$ is less than $0.3\%$ with $L=5$ for the five EcoCAR messages. 
In addition, since the EcoCAR has significantly more traffic than the Toyota Camry, we observe larger $P_e$ for EcoCAR messages than those for the Toyota messages for the same $L$.
We also observe that since we set $\delta$ to a fixed percentage of the message period, messages with larger periods tend to be more advantageous and usually have a smaller $P_e$ with the same $L$. 

\begin{figure}[t!]
\centering
\includegraphics[width=1\columnwidth]{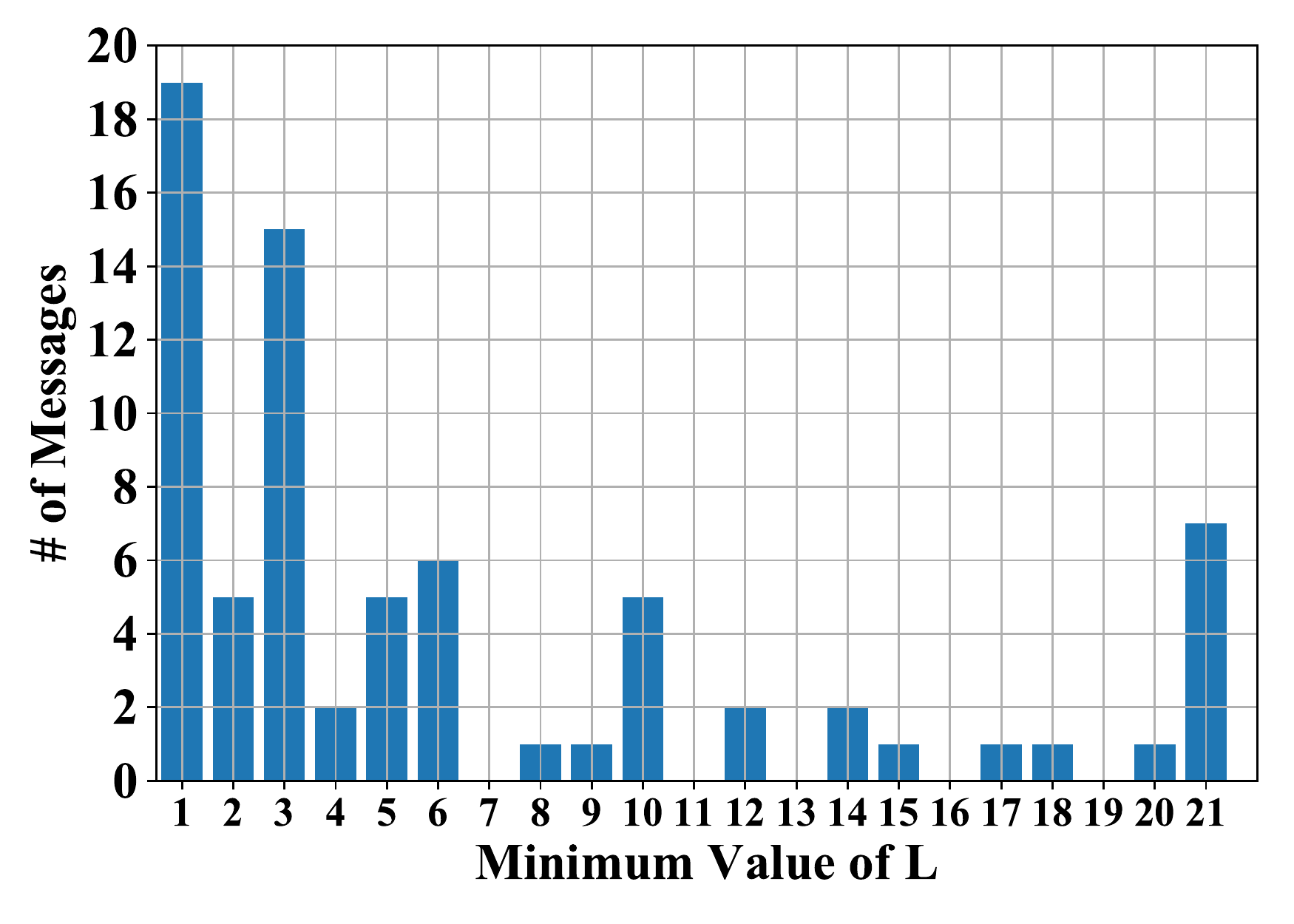}
\caption{Histogram of $L_{\min}$ such that $P_e \leq 1\%$ for EcoCAR messages with $\delta=0.01T$.
}
\label{fig:hist_L_ecocar}
\end{figure}

In order to further understand the minimum $L$ values required by different messages, we choose 85 EcoCAR messages and determine $L_{\min}$ such that the experimental bit error probability is less than or equal to $1\%$ with $\delta = 0.01T$.
Note that we exclude four noisy messages, each of which has a standard deviation of IATs that is $10\%$ larger than its mean IAT. 
As shown in Fig.~\ref{fig:hist_L_ecocar}, a total of 19 messages can use $L=1$.
With $L\leq 5$, a total of 46 messages are eligible, accounting for $51.7\%$ of EcoCAR messages. 
If we set $\delta=0.02T$, a total of 59 messages become eligible, which corresponds to $66.3\%$ of all messages. 

\vspace{0.1cm}
\noindent \textbf{LSB-based covert channel.}
Since the authentication message is hidden in a few bits of the data payload of normal CAN messages in the LSB-based covert channel, the bit error probability will be as low as that of the CAN bus itself. 
According to~\cite{ferreira2004experiment}, the bit error probability of the CAN bus in normal environments (factory production line) and aggressive environments (two meters away from a high-frequency arc-welding machine) are $3.1\times 10^{-7}\%$ and $2.6 \times 10^{-5}\%$, respectively.
In other words, the LSB-based covert channel is very reliable and is not affected by the noise caused by the CAN traffic. 

\subsection{Throughput Performance}
\label{sec:eval:throughput}
In this section, we compute and compare the throughput of the proposed covert channels. 
For ease of discussion, we let the number of bits in the authentication message ($A_m$) be $N_m$, and the number of bits in other fields (SOF, CRC, and EOF) be $N_o$. 
Hence, each authentication frame (denoted as $A_f$) has $N_f = N_m + N_o$ bits. 
However, the number of actually transmitted bits may be more than $N_f$ due to bit stuffing. 
Since the stuffed bit itself may be the first of the five consecutive identical bits, there is at most one stuffing bit every four original bits after the first one. 
Therefore, the actual length of each frame is bounded by
\begin{equation*}
    N'_{f} = N_f + \left\lfloor \frac{N_f - 7 - 1}{4} \right\rfloor, \label{eq:worst_case_frame_size}
\end{equation*}
where $\lfloor \cdot \rfloor$ is the floor function.
Note that we need the $-7$ in the numerator because the EOF is not subject to bit stuffing, and the $-1$ is because the worst-case bit stuffing happens right after the first original bit.  

To quantify the throughput performance, we define two metrics: 1) channel throughput ($R_c$), i.e., the number of bits transmitted per second, and 2) authentication throughput ($R_a$), i.e., the number of authentication bits transmitted per second. 
Since the clock skew is very small and its effect is usually negligible, it is omitted to simplify calculations.

\vspace{0.1cm}
\noindent \textbf{IAT-based covert channel.}
Since every bit is embedded into $L$ consecutive ITTs of $T$ sec, it takes $LT$ time to transmit one bit, which means $R_c = 1/(LT)$. 
The best-case authentication throughput (without bit stuffing) is $R_a=N_m/(N_fLT)$. 
With bit stuffing, $R_a$ may be as low as $N_m/(N'_fLT)$.

\vspace{0.1cm}
\noindent \textbf{LSB-based covert channel.}
Since up to $L$ authentication bits can be hidden in one CAN message, we have $R_c = L/T$. 
Each frame requires $\lceil N_f/L \rceil$ normal CAN messages and a duration of $\lceil N_f/L \rceil T$ to transmit.
Hence, we have $R_c = N_m/(\lceil N_f/L \rceil T) \approx N_mL/(N_fT)$ in the best case and $R_c=N_mL/(N'_fT)$ in the worst case. 

\vspace{0.1cm}
\noindent \textbf{Hybrid covert channel.}
As presented in Section~\ref{sec:hybrid}, the hybrid covert channel splits $A_m$ into two parts with a suitable splitting ratio of $\alpha$ (Eq.~(\ref{eq:splitting_ratio})) and transmits them through the IAT-based and LSB-based covert channels separately. Let $D=\max((\lceil\alpha N_m \rceil + N_o)L_1,  (\lceil(1-\alpha) N_m \rceil + N_o) /L_2)$ be the time it takes to transmit a single frame.
Hence, the channel throughput is $R_c = (N_m+2N_o)/D$. 
The best-case authentication throughput is 
$R_a = N_m/D$.

\vspace{0.1cm}
\noindent \textbf{Throughput comparison.}
In this experiment, we consider a 10-ms CAN message and set $N_m=32$ and $N_o=16$. 
The average throughput is computed for each covert channel for transmitting $1000$ authentication frames. 
Results are provided in Table~\ref{table:throughput_comparison}. 
We can see that $R_c$ of the hybrid covert channel is close to the sum of $R_c$ of individual covert channels, but they are not exactly equal due to bit stuffing.
When $L_1=L_2=1$, the hybrid covert channel has a gain of $49.7\%$ in $R_a$ over individual covert channels.
As either $L_1$ or $L_2$ increases, the throughput of the IAT-based covert channel becomes smaller than that of the LSB-based covert channel, and it is less beneficial to deploy the hybrid covert channel. 
With $L_1=L_2=2$, the hybrid covert channel reduces to the LSB-based covert channel, because no authentication bits are transmitted over the IAT-based covert channel.

\begin{table}[h!]
\caption{Comparison of throughput (in bits per second) of three proposed covert channels with 10-ms CAN messages}
\label{table:throughput_comparison}
\begin{tabular}{|c|c|c||c|c||c|c|}
\hline
\multirow{2}{*}{\textbf{Settings}} & \multicolumn{2}{c||}{\textbf{IAT-Based}} & \multicolumn{2}{c||}{\textbf{LSB-Based}} & \multicolumn{2}{c|}{\textbf{Hybrid}} \\ \cline{2-7} 
 & $R_c$ & $R_a$ & $R_c$ & $R_a$ & $R_c$ & $R_a$ \\ \hline
$L_1=1, L_2=1$ & 100.0  & 61.8 & 100.0  & 61.8 & 198.6  & 92.5 \\ \hline
$L_1=2, L_2=1$ & 50.0 & 30.9 & 100.0 & 61.8 & 148.8 & 69.3 \\ \hline
$L_1=1, L_2=2$ & 100.0 & 61.8  & 197.8 & 122.3 & 296.7 & 138.1 \\ \hline
$L_1=2, L_2=2$ & 50.0  & 30.9 & 197.8 & 122.3 & 197.8 & 122.3 \\ \hline
\end{tabular}
\end{table}

\subsection{Detection of CAN Bus Attacks}
\label{sec:eval:detection}
As described in Section~\ref{sec:proposed_scheme}, TACAN extracts and verifies the authentication message embedded in the IATs or LSBs of normal CAN messages to authenticate the transmitting ECU and also serves the purpose of intrusion detection.
In this section, we consider a TACAN-based detector and evaluate its detection performance against CAN bus attacks.
While we focus on the IAT-based covert channel,  similar results can be found in the other two covert channels.

\vspace{0.1cm}
\noindent \textbf{Detection scheme.}
The TACAN-based detection scheme works as follows: the MN counts the number of consecutive authentication failures.
An authentication failure is either a message reception failure (e.g., lost message, CRC error) or a message verification failure (e.g., invalid counter or digest).
If the counter is larger than or equal to a preset threshold $K$ (e.g., $K=3$), then the MN raises an alert for a CAN bus attack. A successfully received and verified authentication message will reset the detection counter to zero. 



\begin{table}[]
\caption{False alarm probability in the IAT-based covert channel in the normal situation}
\label{table:false_alarm_prob_normal}
\centering
\begin{tabular}{|c|c|c|c|}
\hline
\textbf{Msg ID} & \textbf{$K=1$} & \textbf{$K=2$} & \textbf{$K=3$} \\ \hline
0x0D1 & 0.49\% & 0.19\% & 0.0\% \\ \hline
0x185 & 1.31\% & 0.31\% & 0.0\% \\ \hline
0x22A & 0.0\% & 0.0\% & 0.0\% \\ \hline
0x3FB & 0.64\% & 0.32\% & 0.0\% \\ \hline
0x4D1 & 1.95\% & 0.33\% & 0.0\% \\ \hline
\end{tabular}
\end{table}

\begin{figure}[t!]
    \centering
    \begin{subfigure}[b]{1\columnwidth}
        \includegraphics[trim=0cm 0.5cm 0cm 0cm,clip=true,width=1\textwidth]{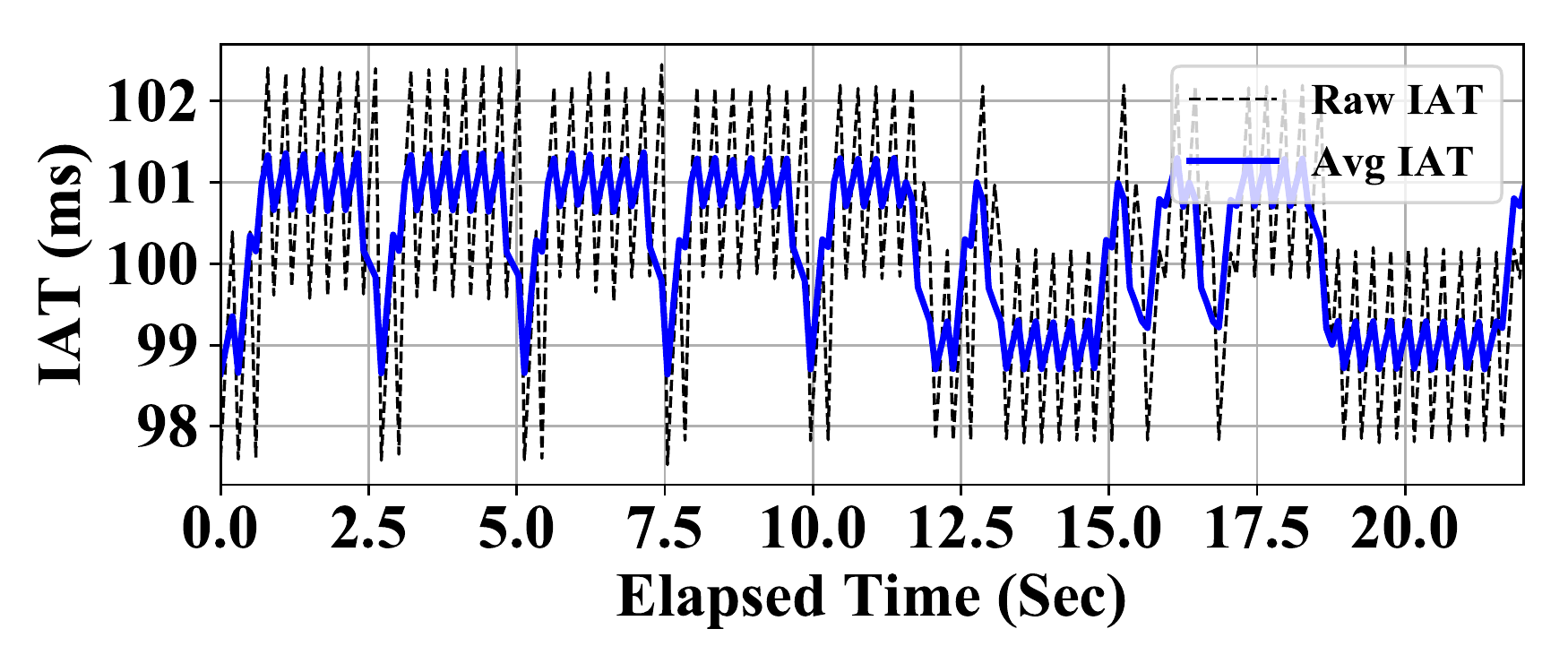}
        \caption{Normal}
        \label{fig:example_of_iat_based_covert_channel_normal}
    \end{subfigure}
    ~
    \begin{subfigure}[b]{1\columnwidth}
        \includegraphics[trim=0cm 0.5cm 0cm 0cm,clip=true,width=1\textwidth]{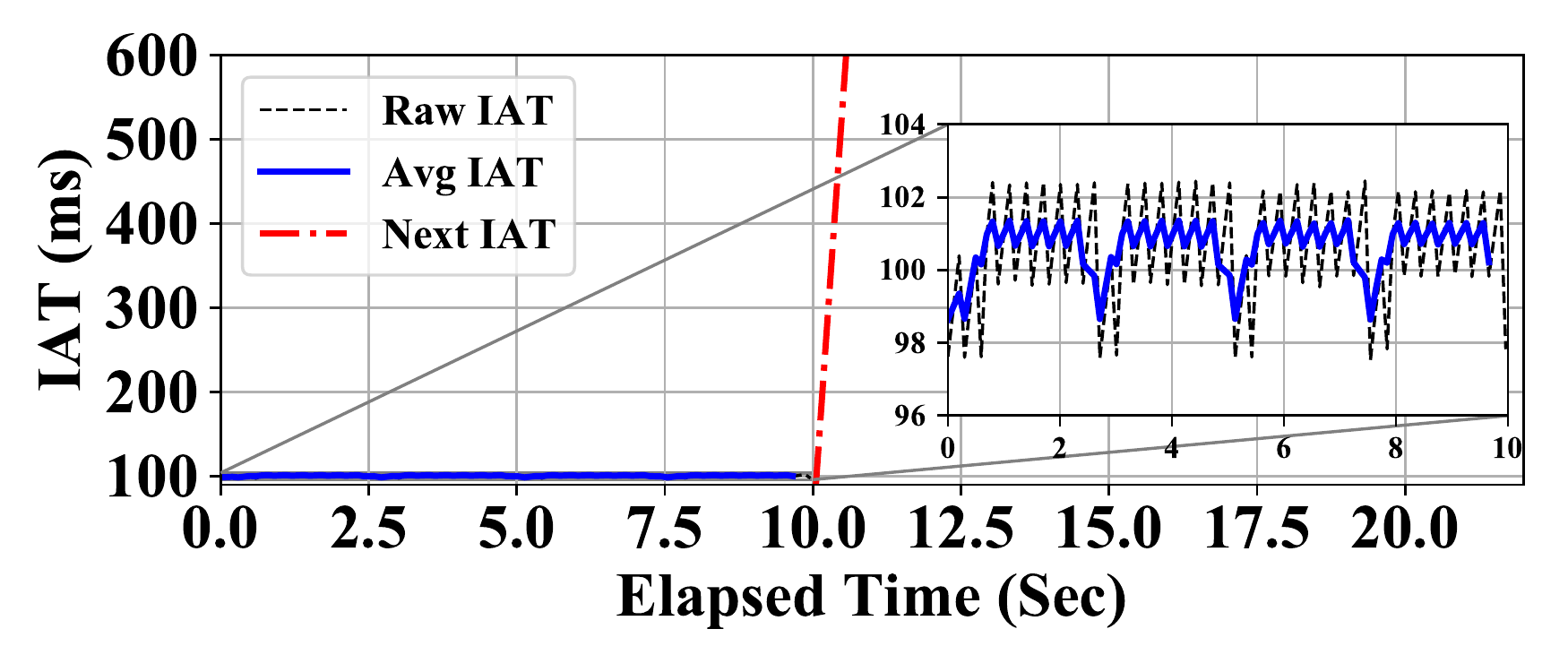}
        \caption{Suspension attack}
        \label{fig:example_of_iat_based_covert_channel_suspension}
    \end{subfigure}
    ~
    \begin{subfigure}[b]{1\columnwidth}
        \includegraphics[trim=0cm 0.5cm 0cm 0cm,clip=true,width=1\textwidth]{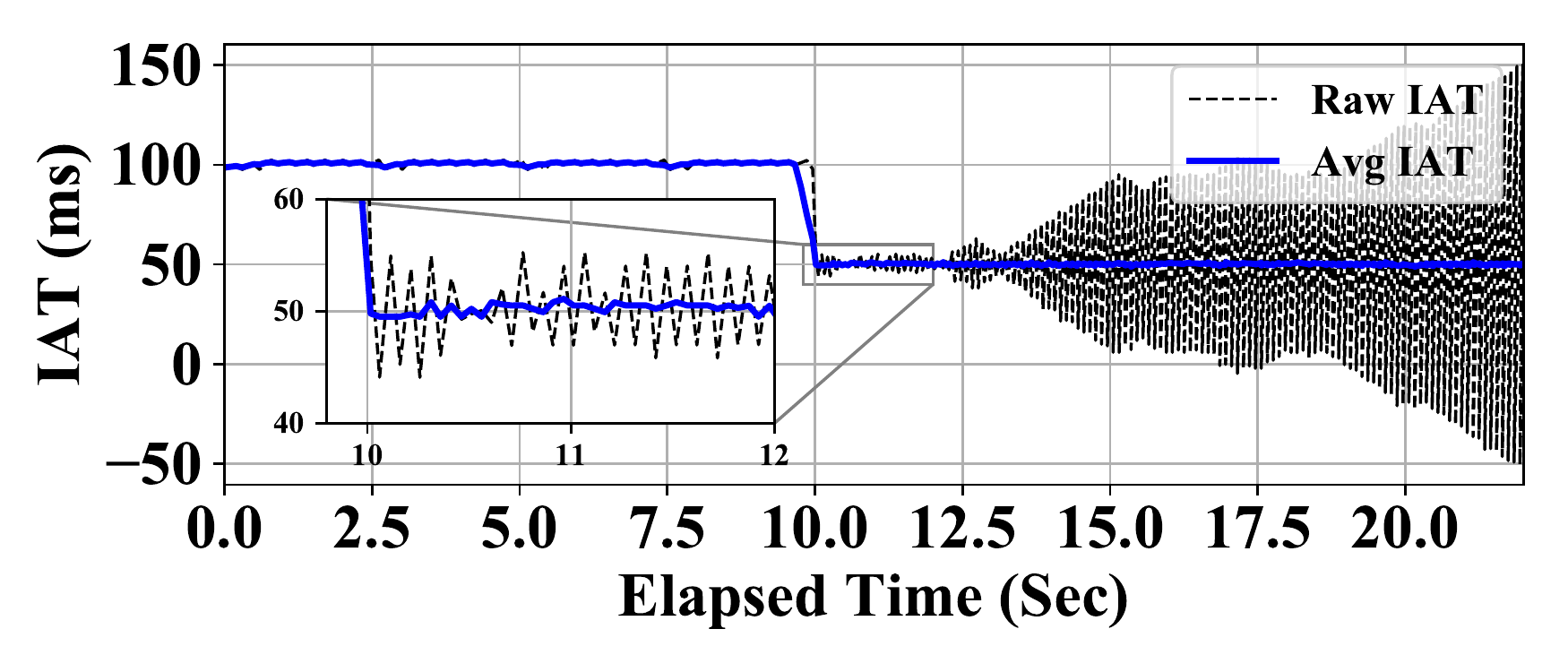}
        \caption{Injection attack}
        \label{fig:example_of_iat_based_covert_channel_injection}
    \end{subfigure}
    \caption{Illustration of the IAT-based covert channel in different situations using message 0x22A from the EcoCAR testbed: (a) normal, (b) under the suspension attack, and (c) under the injection attack.
    In both (b) and (c), the attack starts at around $t=10$ sec. 
    Under the suspension attack, the next message after the attack never arrives, and thus the next expected IAT (in dashed red) will keep increasing. 
    Both suspension and injection attacks will prevent the authentication message from being successfully received by the MN. 
    } 
    \label{fig:example_of_iat_based_covert_channel_under_attack}
\end{figure}

\vspace{0.1cm}
\noindent \textbf{False alarm probability.}
We first discuss the false alarm probability (denoted as $P_{FA}$) of the TACAN-based detector, which is defined as the probability of raising an alarm when there is no attack present. 
We experimentally evaluate the IAT-based covert channel in the normal situation using the five CAN messages from the EcoCAR testbed (see Table~\ref{table:representative_messages}).
An example of the normal IAT-based covert channel transmission is provided in Fig.~\ref{fig:example_of_iat_based_covert_channel_normal}.
In order to minimize bit errors, we set $\delta=0.02T$ and set $L=6$ for the first three messages and set $L=2$ for the last two messages.

As shown in Table~\ref{table:false_alarm_prob_normal}, with a properly configured IAT-based covert channel, we have $P_{FA} \leq 2\%$ with $K=1$. 
When $K=2$, we have $P_{FA} \leq 0.33\%$ for all messages. 
Since none of the messages has three consecutive message reception failures (either a lost message or a CRC error), we have $P_{FA}=0\%$ with $K=3$. 
Note that since the LSB-based covert channel has a very low bit error probability, the probability of a message reception failure in the normal situation is negligible and thus we have $P_{FA}=0\%$.

\vspace{0.1cm}
\noindent \textbf{Detecting suspension and injection attacks.}
As illustrated in Fig.~\ref{fig:example_of_iat_based_covert_channel_suspension}, when the suspension attack happens, the transmission of CAN messages is interrupted, which causes authentication message loss and thus message reception failures. 
Similarly, as shown in Fig.~\ref{fig:example_of_iat_based_covert_channel_injection}, the injection attack also interrupts the transmission of authentication messages, which causes either message loss or CRC errors. 
In this example, we use the 100-ms message with ID=0x139 from the EcoCAR as the attack message.
Since a CAN bus attack like the suspension or injection attack will need to last for a sufficient amount of time to cause enough damage to the CAN bus, it will cause more than $K$ consecutive message reception failures and thus can be detected by the TACAN-based detector with a detection probability of $100\%$. 

\vspace{0.1cm}
\noindent \textbf{Detecting forgery attacks.}
In the forgery attack, the attacker has already compromised an in-vehicle ECU and transmits authentication messages, which can be successfully received by the MN (i.e., CRC check is passed). 
The attacker attempts to generate valid authentication messages which can be verified by the MN to evade the detection of TACAN. 

Since TACAN securely stores both master and session keys in the ECU's TPM to prevent being compromised by the adversary, the attacker has to forge a valid digest for each local counter value without the session key.
With a condensed digest of $M$ bits, the probability of a successful random forgery is $1/2^M$.
For example, when $M=8$, this probability is $1/2^8 \approx 0.4\%$.
In other words, the forgery attack will be detected with a probability of $99.6\%$ with $K=1$.
In general, the probability of $K$ consecutive message verification failures is $(1/2^M)^K$, which decreases as $K$ increases. 
Hence, to guarantee a desired detection probability for a given $K$, a suitable value of $M$ should be chosen.  
Repeated forgeries will be prevented due to the use of monotonic counters. 

\vspace{0.1cm}
\noindent \textbf{Detecting replay attacks.}
A replay attacker has infiltrated the CAN bus and attempts to replay previously transmitted authentication messages of the targeted ECU with the hope of passing the verification process at the MN.
It is easy to see that such attempts will be detected by TACAN with a probability of $100\%$ due to the use of monotonic counters. 

\vspace{0.1cm}
\noindent \textbf{Detecting masquerade attacks.}
As mentioned in Section~\ref{sec:adversary_model}, a masquerade attack (including the more sophisticated cloaking attack \cite{sagong2018cloaking}) requires in-vehicle ECUs to be weakly and/or fully compromised.
As a result, TACAN will force the attacker to perform a forgery or replay attack, not only for the compromised ECU itself, but also for the ECU the attacker attempts to masquerade as. 
Therefore, a masquerade attack will also be detected by TACAN. 

In summary, our proposed TACAN-based detector can identify attacks that interrupt the transmission of normal CAN messages (e.g., suspension and injection attacks), as well as attacks in which attackers fail to generate valid authentication messages (e.g., forgery, replay, and masquerade attacks) with a very high detection probability, while keeping the false alarm probability very low.

\subsection{Accuracy Loss}
\label{sec:eval:accuracy_loss}
In this section, we study the impact of LSB-based covert channel on the accuracy of sensor values. 
We first notice that with changes of a single bit ($L=1$), the accuracy loss has the same scale with the discretization error (resolution) of that sensor value. 
As $L$ increases, the overall throughput will increase (Section~\ref{sec:eval:throughput}), and the accuracy loss will also be larger. 
Hence, manufacturers will need to assess the impact of accuracy loss in CAN data on the functionality and safety, and trade off the accuracy loss against the throughput gain when deploying the LSB-based covert channel.
It is important to highlight that for periodic messages that cannot tolerate accuracy loss, manufacturers may deploy the IAT-based covert channel instead.

\begin{figure}[t!]
    \centering
    \begin{subfigure}[b]{1\columnwidth}
        \includegraphics[trim=-0.5cm 0.7cm 0cm 0cm,clip=true,width=.93\columnwidth]{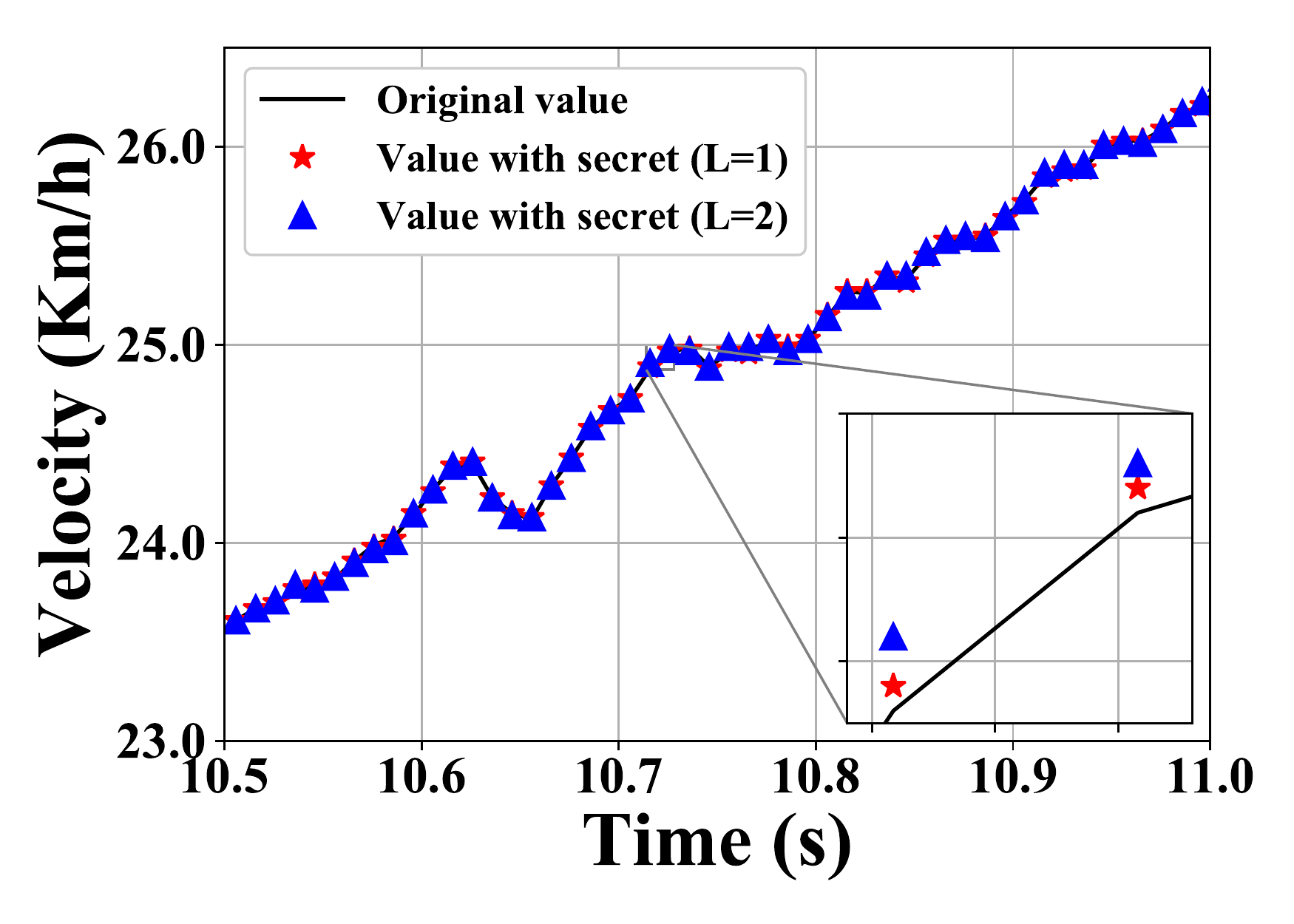}
        \caption{Wheel velocity}
        \label{fig:toyota_lsb_evaluation}
    \end{subfigure}
    ~
    \begin{subfigure}[b]{1\columnwidth}
        \includegraphics[trim=0cm 0.7cm 0cm 0cm,clip=true,width=.9\columnwidth]{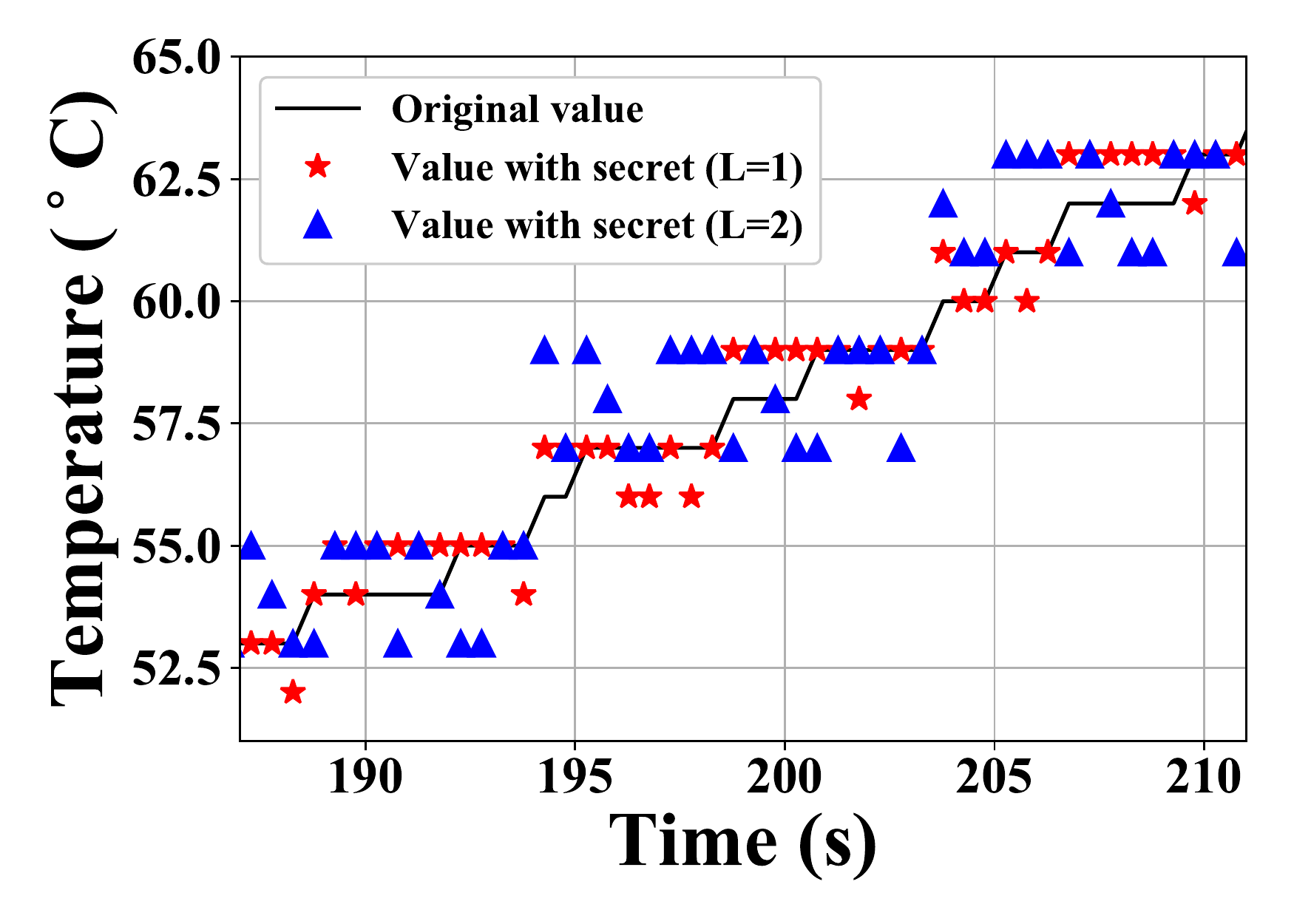}
        \caption{Engine coolant temperature}
        \label{fig:ecocar_lsb_evaluation}
    \end{subfigure}
    \caption{Accuracy loss due to the LSB-based covert channel. (a) Wheel velocity in the Toyota dataset. (b) Engine coolant temperature in the EcoCAR dataset. We observe moderate accuracy loss in both cases. } 
    \label{fig:lsb_evaluation}
\end{figure}

In order to gain a better understanding of the accuracy loss of the LSB-based covert channel, we consider two CAN messages: 1) one that carries wheel velocity values from the Toyota dataset~\cite{toyota2010dataset} and 2) the one carries engine coolant temperature values that we identified through reverse engineering from the EcoCAR dataset~\cite{ecocar}.
In our experiments, we set $L$ to $1$ or $2$ and quantify the accuracy loss in terms of the maximum error, i.e., the maximum deviation from the original values (which are considered as ground truth). 
Note that we intentionally keep $L \leq 2$ in order to avoid significant distortions to the underlying sensor values or jeopardize the functionality of the receiving ECU.

As illustrated in Fig.~\ref{fig:toyota_lsb_evaluation} and highlighted in the magnified box, the maximum error introduced to wheel velocity is $0.01$ km/h for $L=1$ and $0.03$ km/h for $L = 2$, which is considered as insignificant. 
As for the engine coolant temperature in  Fig.~\ref{fig:ecocar_lsb_evaluation}, the maximum error is $1$ $^{\circ}C$ for $L = 1$ and $3$ $^{\circ}C$ for $L = 2$, which are still moderate.
As compared to the median value of $58$ $^{\circ}C$, the errors of $1$ $^{\circ}C$ and $3$ $^{\circ}C$ translate into $1.7\%$ and $5.2\%$ deviations, respectively.
\section{Conclusion}
\label{sec:conclusion}
In this paper, we presented TACAN, a covert channel-based transmitter authentication scheme that allows a MN to verify the authenticity of the transmitting ECU. 
We developed IAT-based, LSB-based, and hybrid covert channels to communicate the authentication information between ECUs and the MN without introducing protocol modifications or traffic overheads. 
We analytically modeled the bit error probability of the IAT-based covert channel as a function of covert channel parameters, and studied its impact on the CAN bus schedulability. 
We experimentally validated and evaluated TACAN and the proposed channels using the EcoCAR testbed and the publicly available Toyota dataset.
Our results show that properly configured IAT-based covert channels have an experimental bit error probabilities of less than 0.3\%. 
We also show that our TACAN-based detector can detect CAN bus attacks with a high detection probability, while keeping the false alarm probability less than 0.33\% by setting the detection threshold to $2$ or higher. 
In addition, we studied the impact of the LSB-based covert channel on sensor values and observed moderate accuracy loss. 








\bibliographystyle{IEEEtran}
\bibliography{./references}

\begin{thebibliography}{10}
\providecommand{\url}[1]{#1}
\csname url@samestyle\endcsname
\providecommand{\newblock}{\relax}
\providecommand{\bibinfo}[2]{#2}
\providecommand{\BIBentrySTDinterwordspacing}{\spaceskip=0pt\relax}
\providecommand{\BIBentryALTinterwordstretchfactor}{4}
\providecommand{\BIBentryALTinterwordspacing}{\spaceskip=\fontdimen2\font plus
\BIBentryALTinterwordstretchfactor\fontdimen3\font minus
  \fontdimen4\font\relax}
\providecommand{\BIBforeignlanguage}[2]{{%
\expandafter\ifx\csname l@#1\endcsname\relax
\typeout{** WARNING: IEEEtran.bst: No hyphenation pattern has been}%
\typeout{** loaded for the language `#1'. Using the pattern for}%
\typeout{** the default language instead.}%
\else
\language=\csname l@#1\endcsname
\fi
#2}}
\providecommand{\BIBdecl}{\relax}
\BIBdecl

\bibitem{ying2019tacan}
X.~Ying, G.~Bernieri, M.~Conti, and R.~Poovendran, ``Tacan: Transmitter
  authentication through covert channels in controller area networks,'' in
  \emph{Proceedings of the 9th ACM/IEEE International Conference on
  Cyber-Physical Systems (ICCPS)}, 2019.

\bibitem{ISO2015CAN}
ISO, \emph{International Standard ISO 11898-1 Road Vehicles-Controller Area
  Network (CAN), Part 1 Data Link Layer and Physical Signaling}, 2015.

\bibitem{bosch1991CAN}
Bosch, ``{CAN} {S}pecification {V}ersion 2.0,'' 1991.

\bibitem{checkoway2011comprehensive}
S.~Checkoway \emph{et~al.}, ``Comprehensive experimental analyses of automotive
  attack surfaces,'' in \emph{Proc. of the 20th USENIX Conf. on Security}, ser.
  SEC'11.\hskip 1em plus 0.5em minus 0.4em\relax Berkeley, CA, USA: USENIX
  Association, 2011, pp. 77--92.

\bibitem{tesla2016remote}
O.~Solon, ``Team of hackers take remote control of {T}esla model {S} from 12
  miles away,'' \url{https://www.theguardian.com}, 2016, accessed: 2018-10-06.

\bibitem{bojarski2016end}
M.~Bojarski \emph{et~al.}, ``End to end learning for self-driving cars,''
  \emph{arXiv preprint arXiv:1604.07316}, 2016.

\bibitem{wyglinski2013security}
A.~M. Wyglinski \emph{et~al.}, ``Security of autonomous systems employing
  embedded computing and sensors,'' \emph{IEEE micro}, vol.~33, no.~1, pp.
  80--86, 2013.

\bibitem{herrewege2011canauth}
A.~Van~Herrewege \emph{et~al.}, ``{CANAuth} -- a simple, backward compatible
  broadcast authentication protocol for {CAN} bus,'' in \emph{ECRYPT Workshop
  on Lightweight Cryptography}, 2011.

\bibitem{hazem2012lcap}
A.~Hazem and H.~Fahmy, ``{LCAP} -- a lightweight {CAN} authentication protocol
  for securing in-vehicle networks,'' in \emph{Proc. of 10th Embedded Security
  in Cars Conference (ESCAR), Berlin, Germany}, vol.~6, 2012.

\bibitem{kurachi2014cacan}
R.~Kurachi \emph{et~al.}, ``{CaCAN}-centralized authentication system in {CAN}
  ({Controller Area Network}),'' in \emph{Proc. of Embedded Security in Cars
  (ESCAR 2014)}, 2014.

\bibitem{radu2016leia}
A.-I. Radu and F.~D. Garcia, ``Leia: A lightweight authentication protocol for
  can,'' in \emph{European Symp. on Research in Computer Security}.\hskip 1em
  plus 0.5em minus 0.4em\relax Springer, 2016, pp. 283--300.

\bibitem{cho2016finger}
K.-T. Cho and K.~G. Shin, ``Fingerprinting electronic control units for vehicle
  intrusion detection,'' in \emph{Proc. of 25th {USENIX} Security Symposium
  ({USENIX} Security 16)}, Austin, TX, 2016, pp. 911--927.

\bibitem{muter2011entropy}
M.~M{\"u}ter and N.~Asaj, ``Entropy-based anomaly detection for in-vehicle
  networks,'' in \emph{2011 IEEE Intelligent Vehicles Symposium (IV)}, June
  2011, pp. 1110--1115.

\bibitem{cho2017viden}
K.-T. Cho and K.~G. Shin, ``Viden: Attacker identification on in-vehicle
  networks,'' in \emph{Proc. of the 2017 ACM SIGSAC Conference on Computer and
  Communications Security}, ser. CCS '17, 2017, pp. 1109--1123.

\bibitem{choi2018voltageids}
W.~Choi \emph{et~al.}, ``{VoltageIDS}: Low-level communication characteristics
  for automotive intrusion detection system,'' \emph{IEEE Trans. Inf. Forensics
  Security}, 2018.

\bibitem{sagong2018cloaking}
S.~U. Sagong, X.~Ying, A.~Clark, L.~Bushnell, and R.~Poovendran, ``Cloaking the
  clock: emulating clock skew in controller area networks,'' in \emph{Proc. of
  the 9th ACM/IEEE Int. Conf. on Cyber-Physical Systems}, ser. ICCPS'18, 2018,
  pp. 32--42.

\bibitem{ying2018shape}
X.~Ying, S.~U. Sagong, A.~Clark, L.~Bushnell, and R.~Poovendran, ``Shape of the
  cloak: Formal analysis of clock skew-based intrusion detection system in
  controller area networks,'' \emph{IEEE Trans. Inf. Forensics Security}, 2019.

\bibitem{murvay2014source}
P.~S. Murvay and B.~Groza, ``Source identification using signal characteristics
  in controller area networks,'' in \emph{IEEE Signal Processing Letters},
  vol.~21, no.~4, April 2014, pp. 395--399.

\bibitem{kneib2018scission}
M.~Kneib and C.~Huth, ``Scission: Signal characteristic-based sender
  identification and intrusion detection in automotive networks,'' in
  \emph{Proc. of the 2018 ACM SIGSAC Conf. on Computer and Communications
  Security}.\hskip 1em plus 0.5em minus 0.4em\relax ACM, 2018, pp. 787--800.

\bibitem{TPMautomotive}
T.~C. Group, ``Tcg tpm 2.0 automotive thin profile. specification version 1.01.
  revision 15,'' Tech. Rep., 2018.

\bibitem{ecocar}
U.~EcoCAR3, ``{University of Washington EcoCAR 3},''
  \url{http://ecocar3.org/washington/about-us/}, 2019, accessed: 2019-1-22.

\bibitem{toyota2010dataset}
J.~Daily, ``Interpreting the {CAN} data for a 2010 {Toyota} {Camry},''
  \url{http://tucrrc.utulsa.edu/ToyotaCAN.html}, 2010, accessed: 2019-1-22.

\bibitem{miller2015remote}
C.~Miller and C.~Valasek, ``Remote exploitation of an unaltered passenger
  vehicle,'' in \emph{Black Hat USA}, 2015.

\bibitem{hoppe2008security}
T.~Hoppe \emph{et~al.}, ``Security threats to automotive {CAN} networks --
  practical examples and selected short-term countermeasures,'' in \emph{Proc.
  of the 27th Int. Conf. on Computer Safety, Reliability, and Security}, ser.
  SAFECOMP '08, 2008, pp. 235--248.

\bibitem{sagong2018exploring}
S.~U. Sagong, X.~Ying, L.~Bushnell, and R.~Poovendran, ``Exploring attack
  surfaces of voltage-based intrusion detection systems in controller area
  networks,'' in \emph{ESCAR Europe 2018}, 2018.

\bibitem{zander2007survey}
S.~Zander \emph{et~al.}, ``A survey of covert channels and countermeasures in
  computer network protocols,'' \emph{IEEE commun. Surveys Tut.}, vol.~9,
  no.~3, pp. 44--57, 2007.

\bibitem{taylor2017enhancing}
J.~M. Taylor \emph{et~al.}, ``Enhancing integrity of modbus tcp through covert
  channels,'' in \emph{Proc. of Inf. Conf. on Signal Process. and Commun. Syst.
  (ICSPCS)}, 2017.

\bibitem{groza2018incanta}
B.~Groza, L.~Popa, and P.-S. Murvay, ``Incanta-intrusion detection in
  controller area networks with time-covert authentication,'' in \emph{Security
  and Safety Interplay of Intelligent Software Systems}.\hskip 1em plus 0.5em
  minus 0.4em\relax Springer, 2018, pp. 94--110.

\bibitem{mills1992NTP}
D.~L. Mills, ``Network time protocol (version 3): {S}pecification,
  {I}mplementation and {A}nalysis,'' RFC 1305, Tech. Rep., 1992.

\bibitem{zander2008measurement}
S.~Zander \emph{et~al.}, ``An improved clock-skew measurement technique for
  revealing hidden services,'' in \emph{Proc. of the 17th Conf. on Security
  Symp.}, 2008, pp. 211--225.

\bibitem{koscher2010experimental}
K.~Koscher \emph{et~al.}, ``Experimental security analysis of a modern
  automobile,'' in \emph{Proc. of the 2010 IEEE Security Privacy}, ser. SP '10,
  2010, pp. 447--462.

\bibitem{lin2012cyber}
C.-W. Lin \emph{et~al.}, ``Cyber-security for the controller area network (can)
  communication protocol,'' in \emph{Proc. of 2012 Int. Conf. on Cyber
  Security}.\hskip 1em plus 0.5em minus 0.4em\relax IEEE, 2012.

\bibitem{cho2016error}
K.-T. Cho and K.~G. Shin, ``Error handling of in-vehicle networks makes them
  vulnerable,'' in \emph{Proc. of the 2016 ACM SIGSAC Conf. on Computer and
  Communications Security}.\hskip 1em plus 0.5em minus 0.4em\relax ACM, 2016,
  pp. 1044--1055.

\bibitem{krawczyk1997hmac}
H.~Krawczyk, M.~Bellare, and R.~Canetti, ``Hmac: Keyed-hashing for message
  authentication,'' Tech. Rep., 1997.

\bibitem{autosar2017crc}
{AUTOSAR}, ``Specification of crc routines autosar cp release 4.3.1,'' Tech.
  Rep., 2017.

\bibitem{szilagy2008flexible}
C.~Szilagy and P.~Koopman, ``A flexible approach to embedded network multicast
  authentication,'' 2008.

\bibitem{goldsmith2005wireless}
A.~Goldsmith, \emph{Wireless communications}.\hskip 1em plus 0.5em minus
  0.4em\relax Cambridge university press, 2005.

\bibitem{davis2007controller}
R.~I. Davis \emph{et~al.}, ``Controller area network (can) schedulability
  analysis: Refuted, revisited and revised,'' \emph{Real-Time Systems},
  vol.~35, no.~3, pp. 239--272, 2007.

\bibitem{groza2012libra}
B.~Groza \emph{et~al.}, ``Libra-can: a lightweight broadcast authentication
  protocol for controller area networks,'' in \emph{Proc. of Int. Conf. on
  Cryptography and Network Security}.\hskip 1em plus 0.5em minus 0.4em\relax
  Springer, 2012, pp. 185--200.

\bibitem{socketCAN}
SocketCAN, ``{Linux-CAN/SocketCAN} user space applictions,''
  \url{https://github.com/linux-can/can-utils}, 2018, accessed: 2018-10-13.

\bibitem{ferreira2004experiment}
J.~Ferreira \emph{et~al.}, ``An experiment to assess bit error rate in {CAN},''
  in \emph{Proc. of 3rd Int. Workshop of Real-Time Networks (RTN2004)}, 2004,
  pp. 15--18.

\end{thebibliography}

\balance
\begin{IEEEbiography}[{\includegraphics[width=1in,height=1.25in,clip,keepaspectratio]{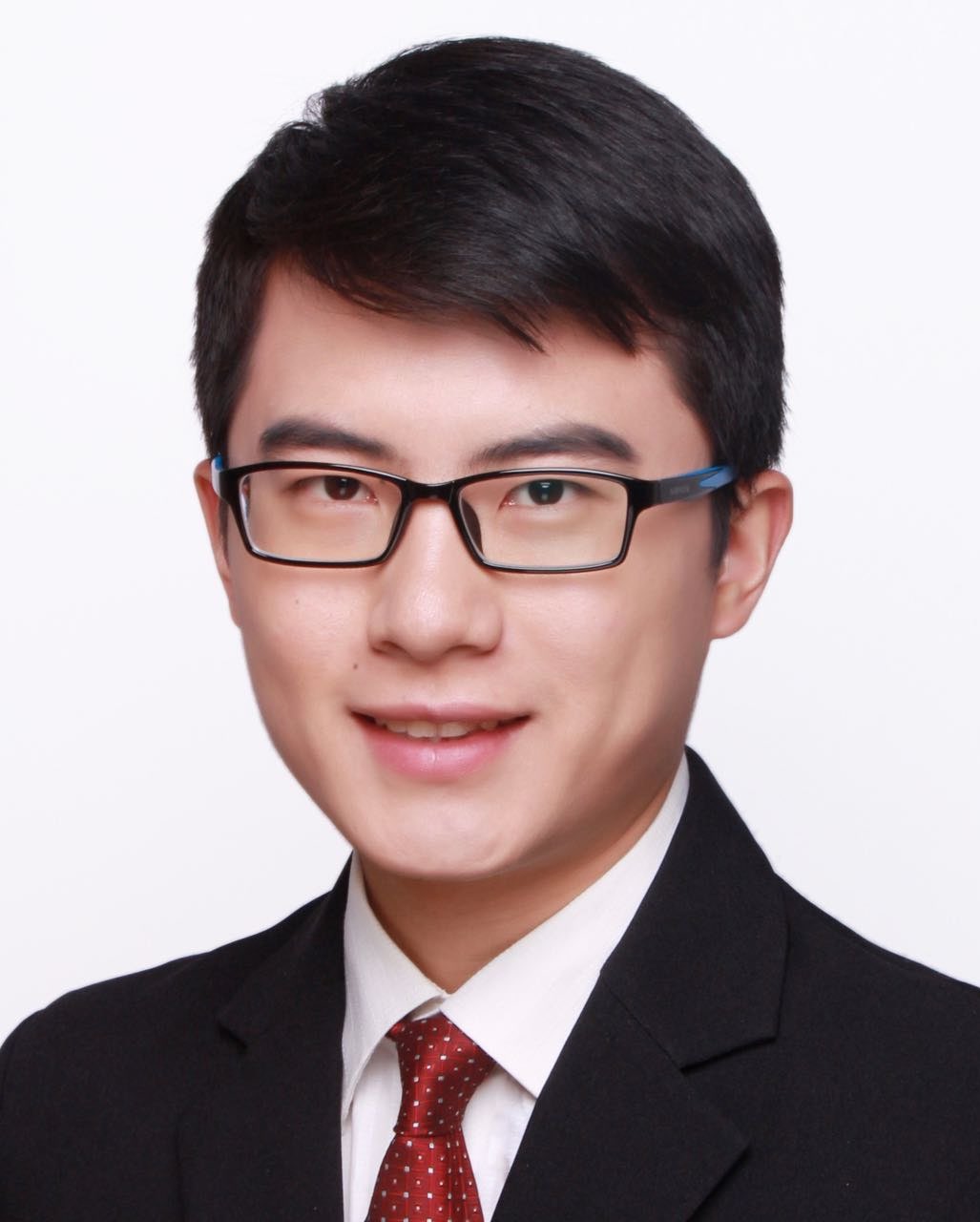}}]{Xuhang Ying}
	(S'15) is a Postdoctoral Research Associate at the Network Security Lab at University of Washington. 
	He received his B.Eng. degree in Information Engineering from the Chinese University of Hong Kong, his M.S. and Ph.D. degrees in Electrical Engineering from the University of Washington in 2013, 2016, and 2018, respectively. 
	His Ph.D. research focused on crowdsensing and resource allocation in shared spectrum. 
	His research interests include Controller Area Networks (CAN) security and wireless security.
\end{IEEEbiography}

\begin{IEEEbiography}[{\includegraphics[width=1in,height=1.25in,clip,keepaspectratio]{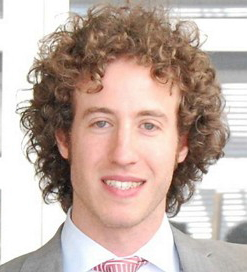}}]{Giuseppe Bernieri}
is a Postdoctoral Researcher at the Department of Mathematics, University of Padua, Padua, Italy. He received the Ph.D. degree in computer science and automation from the University of Roma Tre, Rome, Italy. His research interests include cyber-physical system security, cyber security applied to industrial control systems and critical infrastructures, the development and the implementation of ad-hoc solutions for the detection of cyber-physical threats affecting SCADA systems.
\end{IEEEbiography}

\begin{IEEEbiography}[{\includegraphics[width=1in,height=1.25in,clip,keepaspectratio]{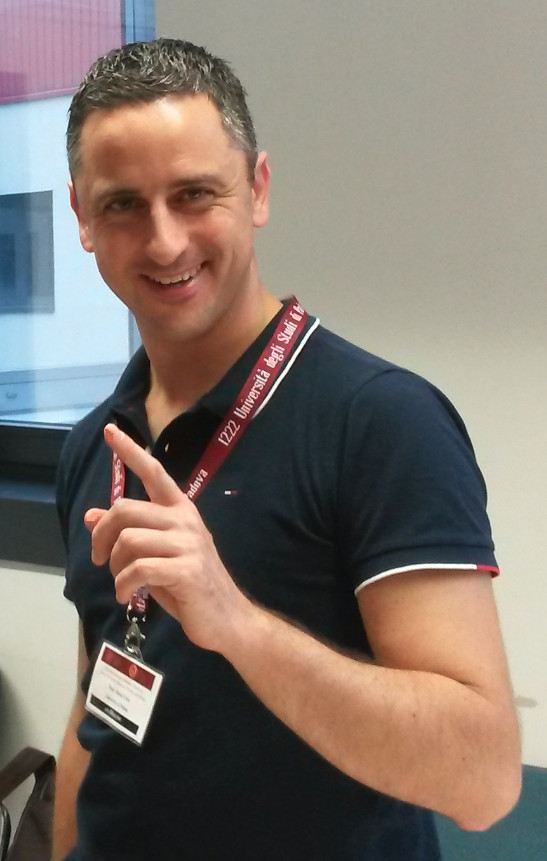}}]{Mauro Conti}
is Full Professor at the University of Padua, Italy. He obtained his Ph.D. from Sapienza University of Rome, Italy, in 2009. After his Ph.D., he was a Post-Doc Researcher at Vrije Universiteit Amsterdam, The Netherlands. In 2011 he joined as Assistant Professor the University of Padua, where he became Associate Professor in 2015, and Full Professor in 2018. He has been Visiting Researcher at GMU (2008, 2016), UCLA (2010), UCI (2012, 2013, 2014, 2017), TU Darmstadt (2013), UF (2015), and FIU (2015, 2016). He has been awarded with a Marie Curie Fellowship (2012) by the European Commission, and with a Fellowship by the German DAAD (2013). His research is also funded by companies, including Cisco and Intel. His main research interest is in the area of security and privacy. In this area, he published more than 200 papers in topmost international peer-reviewed journals and conference. He is Associate Editor for several journals, including IEEE Communications Surveys \& Tutorials, IEEE Transactions on Information Forensics and Security, and IEEE Transactions on Network and Service Management. He was Program Chair for TRUST 2015, ICISS 2016, WiSec 2017, and General Chair for SecureComm 2012 and ACM SACMAT 2013. He is Senior Member of the IEEE.
\end{IEEEbiography}

\begin{IEEEbiography}[{\includegraphics[width=1in,height=1.25in,clip,keepaspectratio]{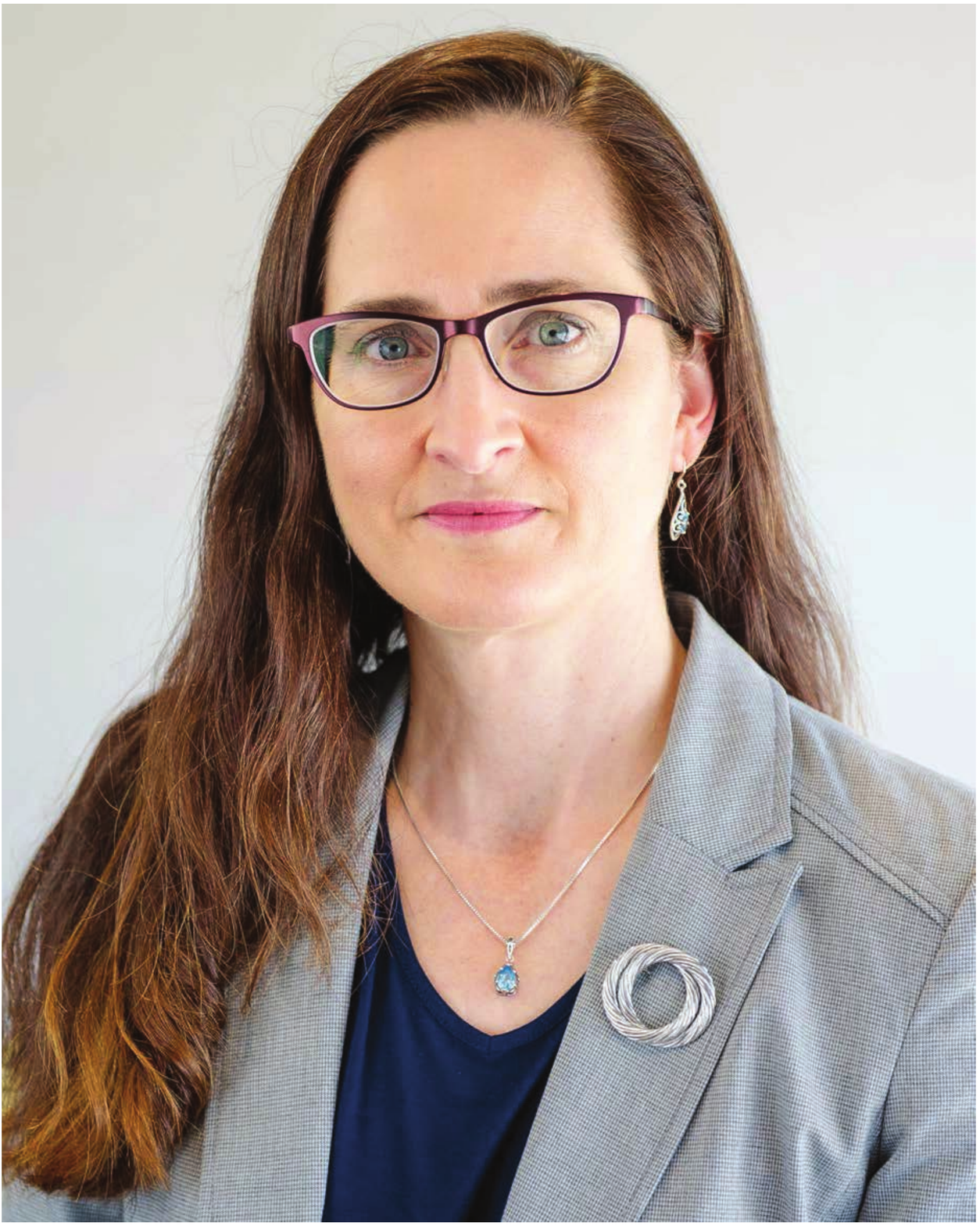}}]{Linda Bushnell}(F'17)
	is a Research Professor and the Director of the Networked Control Systems Lab in the Electrical and Computer Engineering Department at the University of Washington - Seattle. She received the B.S. degree and M.S. degree in Electrical Engineering from the University of Connecticut - Storrs in 1985 and 1987, respectively. She received the M.A. degree in Mathematics and the Ph.D. degree in Electrical Engineering from the University of California - Berkeley in 1989 and 1994, respectively. Her research interests include networked control systems, control of complex networks, and secure-control. She was elected a Fellow of the IEEE for her contributions to networked control systems. She is a recipient of the US Army Superior Civilian Service Award, NSF ADVANCE Fellowship, and IEEE Control Systems Society Distinguished Member Award. She is currently an Associate-Editor for \emph{Automatica} and \emph{IEEE Transactions on Control of Network Systems} and Series Editor for the Springer series \emph{Advanced Textbooks in Control and Signal Processing.} 
	She is currently Chair of the IEEE Control Systems Society (CSS) Women in Control Standing Committee, and Liaison to IEEE Women in Engineering (WIE). She is also the Treasurer of the American Automatic Control  Council (AACC) and a Member of the International Federation of Automatic Control (IFAC) Technical Board.
\end{IEEEbiography}

\begin{IEEEbiography}[{\includegraphics[width=1in,height=1.25in,clip,keepaspectratio]{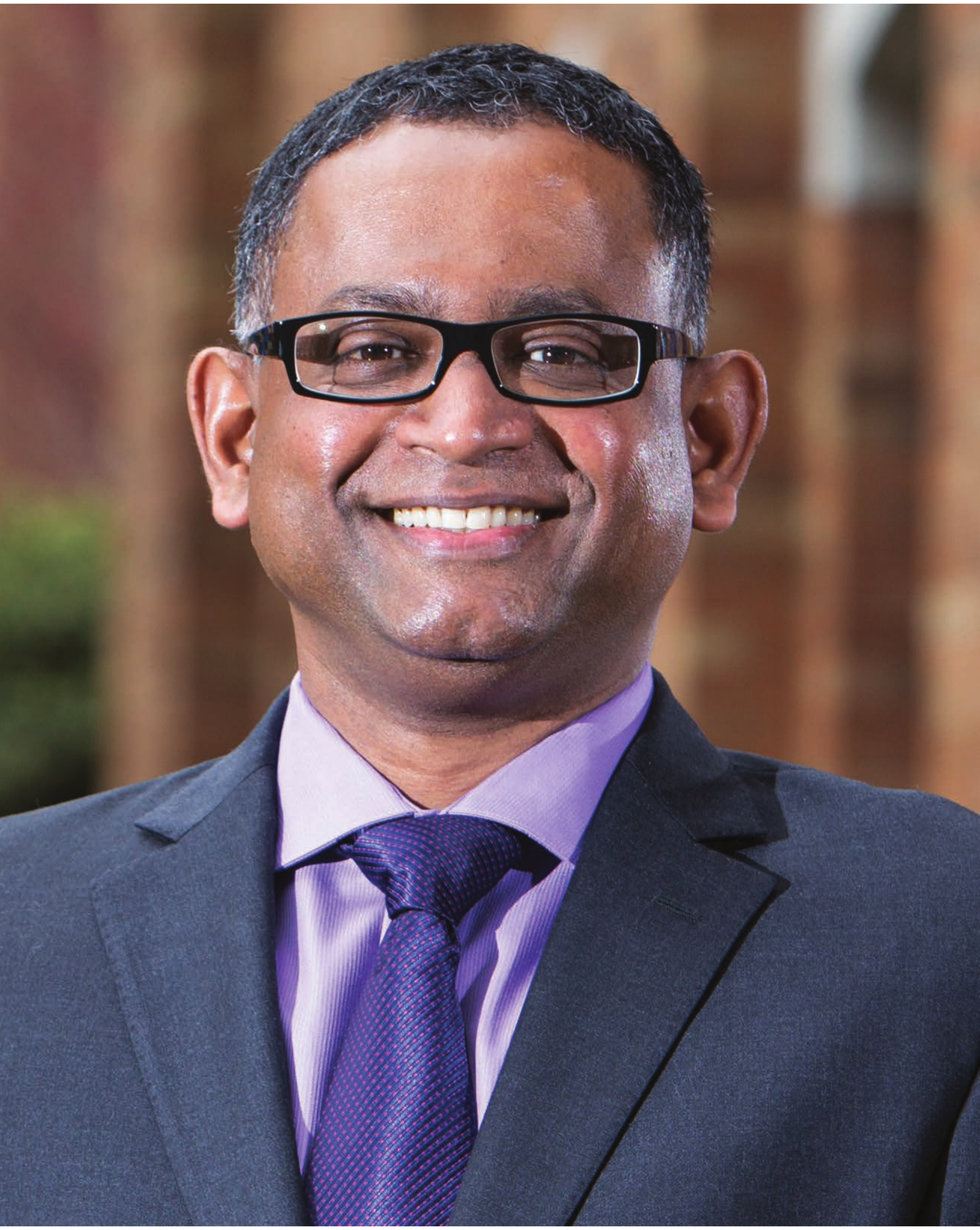}}]{Radha Poovendran}(F'15)
	is a Professor and the Chair of the Electrical and Computer Engineering Department at the University of Washington (UW). He is the Director of the Network Security Lab (NSL) at the University of Washington. He is the Associate Director of Research of the UW Center for Excellence in Information Assurance Research and Education. He received the B.S. degree in Electrical Engineering and the M.S. degree in Electrical and Computer Engineering from the Indian Institute of Technology- Bombay and University of Michigan - Ann Arbor in 1988 and 1992, respectively. He received the Ph.D. degree in Electrical and Computer Engineering from the University of Maryland - College Park in 1999. His research interests are in the areas of wireless and sensor network security, control and security of cyber-physical systems, adversarial modeling, smart connected communities, control-security, games-security and information theoretic security in the context of wireless mobile networks. He is a Fellow of the IEEE for his contributions to security in cyber-physical systems. He is a recipient of the NSA LUCITE Rising Star Award (1999), National Science Foundation CAREER (2001), ARO YIP (2002), ONR YIP (2004), and PECASE (2005) for his research contributions to multi-user wireless security. He is also a recipient of the Outstanding Teaching Award and Outstanding Research Advisor Award from UW EE (2002), Graduate Mentor Award from Office of the Chancellor at University of California - San Diego (2006), and the University of Maryland ECE Distinguished Alumni Award (2016). He was co-author of award-winning papers including IEEE/IFIP William C. Carter Award Paper (2010) and WiOpt Best Paper Award (2012). He holds eight patents in wireless and aviation security.
\end{IEEEbiography}


\end{document}